\documentclass[useAMS,usenatbib]{mn2e}
\usepackage{graphicx}
\usepackage{amsmath}
\usepackage{xcolor}
\usepackage{float}
\usepackage{color}
\usepackage{amssymb}
\usepackage{gensymb}

\newcommand{\be}{\begin{equation}}
\newcommand{\ee}{\end{equation}}

\newcommand{\ifm}[1]{\relax\ifmmode#1\else$\mathsurround=0pt #1$\fi}
\newcommand{\kms}{\ifmmode\,{\rm km}\,{\rm s}^{-1}\else km$\,$s$^{-1}$\fi}

\newcommand{\ltsima}{$\; \buildrel < \over \sim \;$}
\newcommand{\lsim}{\lower.5ex\hbox{\ltsima}}
\newcommand{\gtsima}{$\; \buildrel > \over \sim \;$}
\newcommand{\gsim}{\lower.5ex\hbox{\gtsima}}

\definecolor{green}{rgb}{0,0.5,0}
\definecolor{grey}{rgb}{0.4,0.5,0.7}

\def\M11{M_{11}}
\def\V100{V_{100}}
\def\R1{R_{Mpc}}
\def\T6{T_6}

\begin{document}

\title[Probing the link between quenching and morphological evolution]{Probing the link between quenching and morphological evolution}

\pagerange{\pageref{firstpage}--\pageref{lastpage}} \pubyear{2016}

\author[Koutsouridou and Cattaneo]{I.~Koutsouridou$^{1,2}$, A.~Cattaneo$^{3,4}$
\\
\\
$^1$Dipartimento di Fisica e Astronomia, Universit{\`a} degli Studi di Firenze, Via G. Sansone 1, 50019 Sesto Fiorentino, Italy\\
$^2$INAF/Osservatorio Astrofisico di Arcetri, Largo E. Fermi 5, 50125 Firenze, Italy\\
$^3$Observatoire de Paris, LERMA, PSL University, 61 avenue de l’Observatoire, 75014 Paris, France \\
$^4$Institut d'Astrophysique de Paris, CNRS, 98bis Boulevard Arago, 75014 Paris, France\\
}

\maketitle

\label{firstpage}


\begin{abstract}

We use a semianalytic model of galaxy formation to compare the predictions of two quenching scenarios: halo quenching and black-hole (BH) quenching. After calibrating both models so that they fit the mass function of galaxies, 
BH quenching is in better agreement with the fraction of  passive galaxies as a function of stellar mass $M_*$ and with the galaxy morphological distribution on a star-formation-rate vs. $M_*$ diagram.
Besides this main finding, there are two other results from this research.
First, a successful BH-quenching model requires that minor mergers contribute to the growth of supermassive BHs. If galaxies that reach high $M_*$  through repeated minor mergers are not quenched, there are too many blue galaxies at high masses.
Second, the growth of BHs in mergers must become less efficient at low masses in order to reproduce the $M_{\rm BH}$--$M_*$ relation and the passive fraction as a function of $M_*$, in agreement with the idea that supernovae 
prevent efficient BH growth in systems with low escape speeds. Our findings are consistent with a quasar-feedback scenario in which BHs grow until they are massive enough to blow away the cold gas in their host galaxies and to heat
the hot circumgalactic medium to such high entropy that its cooling time becomes long. They also support the notion that quenching and maintenance correspond to different feedback regimes.

\end{abstract} 

\begin{keywords}
{
galaxies: evolution ---
galaxies: abundances ---
galaxies: star formation
}
\end{keywords}

 
\section{Introduction}
\label{Introduction}

The star-formation (SF) properties of galaxies are closely related to their morphologies \citep{Sandage1986} and display a bimodal distribution \citep{Kauffmann2003,Baldry2004}.
Four scenarios have been proposed to explain these observations:
\begin{enumerate}
\item Elliptical galaxies are passive because they converted all their gas into stars very rapidly at high redshift \citep{Sandage1986}. Coupled to the hypothesis that elliptical galaxies
formed through mergers \citep{Toomre1972} and to the observation that mergers can drive starbursts \citep{Sanders1988}, this picture explains why rapid depletion of gas is associated with elliptical morphologies. Its main problem is that gas-rich mergers leave behind persistent blue cores \citep{Cattaneo2005,Springel2005a}. Elliptical galaxies with blue cores (E+A galaxies) are observed, but they are exceedingly rare.
\item Central black holes (BHs) grow until they are massive enough to heat and/or blow away all the gas in their host systems (BH quenching; \citealp{Silk1998,Fabian1999,King2003,Chen2020}).
In a version of this scenario (\citealp{Springel2005b,Hopkins2006,
Hopkins2008a}), gas-rich mergers feed the growth of supermassive BHs until quasar feedback blows away all the leftover gas.
\item The transition from star-forming (blue) to passive (red) galaxies is linked to the disruption of cold filamentary flows and the appearance of shock-heated gas in massive systems (halo quenching; \citealp{Dekel2006}). 
This picture explains the galaxy colour--magnitude bimodality in quantitative detail \citep{Cattaneo2006}. The link to morphology is less direct (the shutdown of gas accretion prevents merger remnants from regrowing discs).
\item This scenario can be characterised as BH or halo quenching followed by BH maintenance.
All the above scenarios share a common weakness: they do not explain what prevents the accretion of gas from restarting after the initial shutdown of SF (the so-called maintenance problem). 
Model (iv) is model (ii) or (iii) with the addition of  mechanical heating by radio galaxies, which solves the maintenance problem by preventing the hot gas from cooling \citep{Bower2006,Cattaneo2006,Croton2006,
Somerville2008}. In this picture, BHs constitute the link between quenching and morphology. Elliptical galaxies are passive because they are those with the most massive BHs.
\end{enumerate}

The problem is complicated because all the mechanisms above play a role to some extent. The question is which ones dominate. An observational approach to it is to ask which observable is the best predictor of quiescence. \citet{Cheung2012, Fang2013} and \citet{Barro2017} showed that the surface density in the central kiloparsec, $\Sigma_{\rm 1}$, is the best predictor of quiescence in large surveys (galaxies are quiescent above a critical $\Sigma_{\rm 1}$ that depends on $M_*$). \citet{Fang2013} also mentioned a good correlation with the central stellar velocity dispersion $\sigma_{\rm c}$ and noted that this would follow naturally from that with $\Sigma_{\rm 1}$ since $\sigma_{\rm c}$ and $\Sigma_{\rm 1}$ are closely related. \citet{Terrazas2016} used observations of galaxies with direct measurements of the BH mass $M_{\rm BH}$ and showed that, for a same $M_*$, passive galaxies have higher $M_{\rm BH}$ than star-forming galaxies. \citet{Bluck2020} used the empirical $M_{\rm BH}$--$\sigma_{\rm c}$ relation \citep{Ferrarese2000, Gebhardt2000,Saglia2016, Sahu2019} to conclude that quiescence correlates with $M_{\rm BH}$  more strongly that it does with stellar mass, halo mass, or any other observable property. An equally strong correlation with $\Sigma_{\rm 1}$ would not modify this conclusion to the extent that $\Sigma_{\rm 1}$ and $\sigma_{\rm c}$ are both tracers of $M_{\rm BH}$ (see \citealp{Sahu2022} for an observational analysis of the $M_{\rm BH} - \Sigma_{\rm 1}$ relation).

In this article, we use a semianalytic model (SAM) of galaxy formation to shed some more light on this topic and compare the observational implications of {\it BH quenching} vs. 
{\it halo quenching} for:

a) the galaxy stellar mass function (GSMF), 

b) the fraction of passive galaxies as a function of $M_*$, 

c) the $M_{\rm BH}$--$M_*$ relation, 

d) galactic morphologies.

\noindent  Point (d) is important to break potential degeneracies of the SAM. A model with an over-propensity for early-type morphologies could reproduce the GSMF and the passive fraction correctly if this over-propensity were compensated by lower BH accretion rates, so that the product $M_{\rm BH}/M_{\rm bulge}\times M_{\rm bulge}/M_*$ stays the same (here $M_{\rm bulge}$ is the stellar mass of the bulge).

The article starts by presenting the SAM used for this study (Section~2), which we call {\sc GalICS}~2.2, since it is an updated version of 
{\sc GalICS}~2.1 \citep{Cattaneo2020}.
{\sc GalICS}~2.2 contains two options for the quenching model: halo quenching (model~A) and BH quenching (model~B).
It also includes two models for the morphologies of merger remnants.
In model~1 (also referred to as the threshold model), there is a sharp separation between major and minor mergers.
Model~2 follows numerical simulations  by \citet{Kannan2015} and assumes a smoother transition.
In Section~3, we compare the results of the four combinations A1, A2, B1, B2.
In Section~4, we summarize and discuss the conclusions of the article.

\section{The model}

This section describes how {\sc GalICS}~2.2 follows the hierarchical growth of dark-matter (DM) haloes (Section~2.1), the accretion of gas onto haloes and galaxies (Section~2.2),
star formation (Section~2.3), stellar feedback (Section~2.4), disc instabilities (Section~2.5),  mergers (Section~2.6), the growth of supermassive BHs (Section~2.7), quenching (Section~2.8) and ram-pressure stripping  (Section~2.9).
The presentation is more synthetic in Sections~2.1 to 2.4, where the differences with {\sc GalICS}~2.1 \citep{Cattaneo2020} are small (or inexistent, in Sections~2.1--2.2), and more detailed in Sections~2.5 to 2.9,
where the new developments are.

\subsection{\textit{N}-body simulation}

{\sc GalICS}~2.2 uses the  same DM merger trees as {\sc GalICS}~2.1. They were extracted from a cosmological N-body simulation with $\Omega_{\rm M} = 0.308, \Omega_{\rm \Lambda} = 0.692, \Omega_{\rm b} = 0.0481$ and $\sigma_8 = 0.807$ \citep{planck_etal14}. The simulation has a volume of $(100\,{\rm Mpc})^3$ and contains $1024^3$ particles, so that the halo mass functions are reproduced down to $M_{\rm vir}\simeq 4 \times 10^9\,{\rm M_\odot}$.  
It has been evolved from $z=16.7$ to $z=0$ storing 287 outputs equally spaced in the logarithm of the expansion factor.

The virial mass $M_{\rm vir}$, the virial radius $r_{\rm vir}$ and the virial angular momentum $\bf{\rm J}_{\rm vir}$ are computed at each timestep for each halo containing at least 100 particles, as in \citet{Cattaneo2017}. 
The virial masses, orbits and merging times of ghost subhaloes, i.e. subhaloes that fall below the resolution limit of the simulation,  are computed as in \citet{Cattaneo2020}.

\subsection{Gas accretion}
\label{Gas accretion}

Gas accretion contains two separate aspects: the accretion of gas onto haloes and its later accretion (or reaccretion) onto galaxies. Let us start from the former.
The gaseous mass that accretes onto a halo at each timestep is computed by requiring that:
\begin{equation}
\label{e:accr}
M_{\rm bar} = f_{\rm b}{M}_{\rm vir},
\end{equation}
where $M_{\rm bar}$ is the total baryonic mass that has accreted onto the halo (inclusive of any ejected material) and $f_{\rm b}$ is the halo baryon fraction (see below).
If the left-hand side of Eq.~(\ref{e:accr}) is smaller than the right-hand side, then gas accretes onto the halo until the validity of Eq.~(\ref{e:accr}) is restored
(see Fig.~1 of \citealp{Cattaneo2020} for the details of how this accretion scheme is implemented).

In massive haloes, $f_{\rm b}$ should approach the universal baryon fraction $\Omega_{\rm b}/\Omega_{\rm M}$. Photoionization heating (e.g. \citealp{Efstathiou1992,Gnedin2000}) and stellar feedback (e.g. \citealp{Brook2012, Wang2017, Tollet2019}), however, suppress gas accretion onto low-mass haloes, so that their baryon fraction is lower than the universal one.
We follow \citet{Cattaneo2020} and assume that:
\begin{equation}
f_{\rm b}(\upsilon_{\rm vir}) = \frac{\Omega_{\rm b}}{\Omega_{\rm M}} \bigg[{\rm erf}\Big( \frac{\upsilon_{\rm vir}}{\sigma} \Big) - \frac{2}{\sqrt{\pi}} \frac{\upsilon_{\rm vir}}{\sigma} e^{-\left(\frac{\upsilon_{\rm vir}}{\sigma}\right)^2} \bigg],
\end{equation}
where $v_{\rm vir}=\sqrt{{\rm G}M_{\rm vir}/r_{\rm vir}}$ is the virial velocity of the halo and
$\sigma = 34\,{\rm km \, s^{-1}}$ is a parameter of the SAM that we fix in agreement with the results of the {\sc NIHAO} simulations \citep{Tollet2019}.

The gas that accretes onto a halo can stream onto the central galaxy in freefall (cold-mode accretion) or it may be shock-heated and become part of a hot atmosphere. The choice is based on a shock-stability criterion that compares the compression time with the post-shock cooling time on a halo-by-halo basis. 
This criterion, which improves previous work by \citet{Dekel2006}, has been discussed at length in \citet{Cattaneo2020} and \citet{Tollet2022}, who have performed a detailed comparison with cosmological zoom simulations. Hence, there is no need for us to elaborate on it here.
Our SAM also considers the effects of the Kelvin-Helmholtz instability in haloes where cold filamentary flows and a hot atmosphere coexist.
The Kelvin-Helmholtz instability contributes to the disruption of the filaments in such situations (\citealp{Cattaneo2020} for details).

Shock-heated gas can accrete onto the central galaxy by cooling (hot-mode accretion), which we compute with the standard method of \citet{White1991}. 
The cooling rate of the hot gas depends on its density distribution $\rho_{\rm h}(r)$.
As in \citet{Cattaneo2020}, we assume a density profile of the form:
\begin{equation}
\label{Faltenbacher}
\rho_{\rm h}(r) = \rho_{\rm h,0} \bigg( 1+ \frac{r}{\alpha_{\rm g}r_0} \bigg)^{-3},
\end{equation}
where $r_0$ is the core radius of the DM halo, fitted with a \citet*[NFW]{Navarro1997} profile, and $\rho_{{\rm h},0}$ is the central density of the hot gas, determined by  the condition that $\rho_{\rm h}/\rho_{\rm NFW} \rightarrow M_{\rm hot}/M_{\rm vir}$ for $r \rightarrow \infty$
($M_{\rm hot}$ is the mass of the hot gas within the halo). We set $\alpha_{\rm g}=0.59$ in agreement with adiabatic 
cosmological hydrodynamic simulations by \citet{Faltenbacher2007}.

\begin{figure}
\begin{center}
\includegraphics[width=1.\hsize]{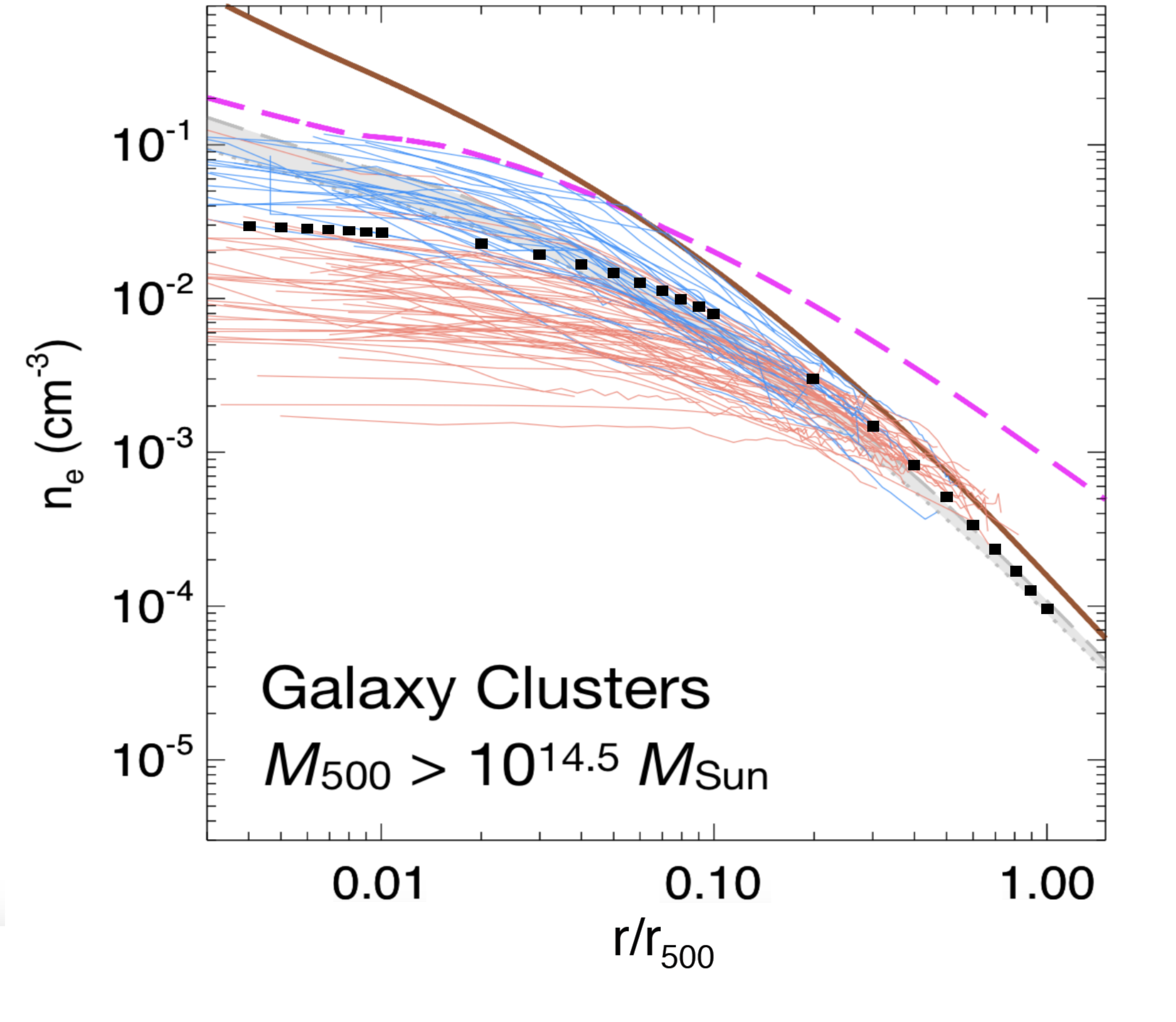} 
\end{center}
\caption{The hot-gas density profile of the biggest cluster in our computational volume ($M_{500}=10^{14.6}\:{\rm M_\odot}$) as predicted by Eq.~(\ref{Faltenbacher}) (black squares), overplotted on  Fig.~2 of \citet{Voit2019c}. The blue  and red curves show the density profiles of cool-core and non-cool-core clusters in the ACCEPT sample \citep{Cavagnolo2009}, respectively. The brown solid curve shows the NFW density profile rescaled by the universal baryon fraction. The magenta dashed curve shows the precipitation limit.
For densities above this limit, the cooling time is shorter than ten freefall times and the gas starts fragmenting into clouds \citep{Voit2019c}. 
The gray shaded area shows hybrid profiles obtained by \citet{Voit2019b} combining the cosmological and precipitation limits (that is, the constraints from the brown solid curve and the magenta dashed curve, respectively). 
The comparison with \citet{Voit2019b} is a minor point in relation to the goals of this article. Interestingly, however, there is a very good agreement between his model and ours at $r\gsim 0.05\,r_{500}$.
At $r< 0.05\,r_{500}$, the gray shaded area is not only above the black squares but also close to the upper boundary of the observationally permitted region. Not all clusters are close to the precipitation limit, but most cool-core clusters (blue curves) are.
}
\label{i:Voit}
\end{figure}

The black squares in Fig.~\ref{i:Voit} show the electron-density distribution $n_{\rm e}={14\over 16}{\rho_{\rm h}\over m_{\rm p}}$ predicted by Eq.~(\ref{Faltenbacher})
for a cluster halo, in which the gas is approximately adiabatic ($m_{\rm p}$ is the proton mass; the $14/16$ coefficient is for  a plasma that is three quarter hydrogen and one quarter helium in mass).
The blue and the red curves are observational data  for cool-core  and non-cool-core clusters in the same range of masses, respectively.
Real clusters have central electron densities that span two orders of magnitude.
Denser cores correspond to lower central entropies.
Cool-core clusters are those in which the hot X-ray gas has cooled and condensed into the central region.
The lowest central densities occur where heating has prevailed over cooling, so that the central entropy has gone up.
In our model, neither cooling nor any non-gravitational heating mechanism, such as feedback from active galactic nuclei (AGN), have any effect on $n_{\rm e}(r)$.
Hence, it is remarkable that our predictions (the black squares) lie right at the boundary between the density profiles of cool-core and non-cool-core clusters.

We model a galaxy as the sum of four components: a disc, a bar/pseudobulge, a classical bulge, and a central cusp, composed of the gas that falls to the centre and undergoes starbursts during mergers \citep{Cattaneo2017}.
We call it cusp because this material is responsible for the central light excesses in the photometric profiles of cuspy ellipticals, while the rest of the bulge is composed of the pre-existing stellar population
 \citep{Kormendy2009}.

All the gas that accretes onto galaxies is added to the disc component, the radius of which is computed assuming the conservation of angular momentum \citep{Mo1998}.
This assumption may be inaccurate \citep{Jiang2019} but is highly standard (see \citealp{Knebe2015} and \citealp{Somerville2015a} for on overview of SAMs) and gives mean
disc sizes in good agreement with the observations, even though the scatter around the mean is larger in the SAM than in the observations because of the large intrinsic scatter in the halo spin parameter
\citep{Cattaneo2017}.

\subsection{Star formation and chemical enrichment}
Star formation, stellar evolution, chemical evolution and feedback are computed separately for each component. In {\sc GalICS},
there are two modes of star formation: a quiescent mode, associated with the disc, and a starburst mode, associated with cusps \citep{Hatton2003,Cattaneo2006,Cattaneo2017}.
There is no SF in bulges, where stellar mass loss is the only source of gas.
Bars are the most difficult component to fit in this schematic picture. We make the simplifying assumption of treating their SF as quiescent.

We define the SF timescale of a component as the time in which it would consume its whole gas reservoir, were its SF rate (SFR) constant.
Many authors call this quantity the gas-depletion timescale. We do not follow this convention here  because SF is not the only process that depletes galaxies of gas. Feedback, too, plays a role.

\citet{Bigiel2008,Bigiel2011} advocated a constant SF timescale of $\sim 2\,$Gyr in disc galaxies. Their study suggests that a similar conclusion applies to starbursts with a timescale $\sim 10$ times shorter.
More complex models have been proposed based on both observations \citep{Kennicutt1998,Kennicutt2012,Kennicutt2021} and theory \citep{Krumholz2012,Somerville2015b}.
We ourselves had considered a SF timescale proportional  to the dynamical time in our previous work \citep{Hatton2003,Cattaneo2006,Cattaneo2017}. However,
assuming a constant SF timescale has the practical advantage that our SFRs do not depend on the sizes of discs.
This prevents the possibility of artificial starbursts in a tail of discs with abnormally small exponential scale-lengths (Section~2.2, final paragraph).

In this article, we have treated the SF timescales for discs and starbursts as free parameters of the SAM.
We have found good fits to the observations using $1\,$Gyr for the disc SF timescale and $0.2\,$Gyr for the cusp SF timescale.
The timescale for discs is shorter than the one found by \citet{Bigiel2011} but in agreement with \citet{Combes2013}.

 Chemical evolution is modelled as in \citet{Koutsouridou2019}. The only novelty is that we no longer assume primordial metallicity for the intergalactic medium.
The enrichment of  the intergalactic medium is computed self-consistently as in \citet{Cattaneo2020}.
 
\subsection{Stellar feedback}
\label{Stellar feedbac}
In \citet{Cattaneo2020}, we introduced a new model for stellar feedback that takes into account the multiphase structure of galactic winds by separating them into a cold component with mass-loading factor $\eta_{\rm c}$ and a hot component with mass-loading factor $\eta_{\rm h}$, so that 
\begin{equation}
\eta\equiv\frac{\dot{M}_{\rm out}}{\rm SFR} = \eta_{\rm c} + \eta_{\rm h} = 2 \bigg(\frac{\epsilon_{\rm SN_{\rm c}}}{v_{\rm w_{\rm c}}^2} +\frac{\epsilon_{\rm SN_{\rm h}}}{v_{\rm w_{\rm h}}^2}\bigg) E_{\rm SN} \Psi_{\rm SN}.
\label{e:eta}
\end{equation}
Here $\dot{M}_{\rm out}$ is the total mass outflow rate; $\Psi_{\rm SN}=1/140$ is the number of supernovae per unit stellar mass formed for a \citet{Chabrier2003} initial mass function; $E_{\rm SN} = 10^{51}\,{\rm erg}$ is the energy released by one supernova; $\epsilon_{\rm SN_{c}}$ and $\epsilon_{\rm SN_{h}}$ 
are the fractions of this energy used to power the cold and the hot component of supernova-driven winds, respectively; $v_{\rm w_{c}}$ and $v_{\rm w_{h}}$ are the respective speeds.

\citet{Cattaneo2020} modelled the fraction of the total energy from supernovae deposited into the hot component with the functional form:
\begin{equation}
\epsilon_{\rm SN_{\rm h}} = {\rm min} \bigg[ \bigg(\frac{v_{\rm vir}}{v_{\rm SN}} \bigg)^{\alpha_v}  (1+z)^{\alpha_{z}},1\bigg]
\label{e:e_SN}
\end{equation}
based on the results of cosmological hydrodynamic zoom simulations \citep[NIHAO]{Tollet2019}, where $\alpha_{v}$, $\alpha_{z}$ and $v_{\rm SN}$ 
are free parameters of the SAM; $v_{\rm SN}$  is the virial velocity for which $\epsilon_{\rm SN_{\rm h}}=1$ at $z=0$.
The same model applies here with $\alpha_{v}=-3.6$, $\alpha_{z}=2.5$ and $v_{\rm SN}=55\,{\rm km/s}$ (see Table~1 of \citealp{Cattaneo2020} for the values of the same parameters in {\sc GalICS}~2.1).
The hot wind is assumed to flow out at the escape speed  ($v_{\rm w_{h}}=v_{\rm esc}$).

The assumptions in the paragraph above are used to compute $\eta_{\rm h}$; $\eta_{\rm c}$ is determined by the fountain fraction:
\begin{equation}
\epsilon_{\rm fount} \equiv {\eta_{\rm c}\over \eta_{\rm c}+\eta_{\rm h}},
\end{equation}
since it is the cold wind that falls back and forms the galactic fountain. In {\sc GalICS}~2.1, $\epsilon_{\rm fount}$ was a free parameter of the SAM and its value was the same for all galaxies ($\epsilon_{\rm fount}=0.7$).
However, cosmological hydrodynamic
simulations show that $\epsilon_{\rm fount}$ decreases with increasing $v_{\rm vir}$ (Fig.~10
of \citealp{Christensen2016} and Fig.~12 \citealp{Tollet2019}). Here we assume:
\begin{equation}
\epsilon_{\rm fount} = 0.7\,{\rm min} \bigg[\frac{\upsilon_{\rm break}}{\upsilon_{\rm vir}},1\bigg],
\end{equation}
where $v_{\rm break}$ is the virial velocity below which $\epsilon_{\rm SN_{\rm h}}=1$ in Eq.~(\ref{e:e_SN}).

Our model for what happens to the ejected gas is the same as in {\sc GalICS}~2.1.
The reaccretion timescale for the cold gas in the galactic fountain is computed with Eq.~(35) of \citet{Cattaneo2020}. The fountain has a shorter thermal-evaporation timescale when the virial temperature is higher (\citealp{Cattaneo2020}, Eq.~36).
Hot winds escape from the halo entraining the hot circumgalactic medium with them if their momentum is
$\dot{M}_{\rm out}^{\rm hot}v_{\rm esc} > M_{\rm hot} g_{\rm h}$
where $\dot{M}_{\rm out}^{\rm hot}$ is the outflow rate of the hot component,
$M_{\rm hot}$ is the mass of the hot circumgalactic medium (i.e. the hot gas within the halo), and $g_{\rm h}$ is the gravitational acceleration at the center of the DM halo.
They mix with the hot circumgalactic medium if this condition is not satisfied.

\subsection{Disc instabilities}
\label{Disc instabilites}
Since \citet{Mo1998} and \citet{Bosch1998}, all SAMs of disc instabilities have been based on the   
\citet*[ELN]{Efstathiou1982}'s
criterion ({\sc Galacticus}: \citealp{Benson2012}; {\sc Morgana}: \citealp{LoFaro2009}; {\sc Sag}: \citealp{Gargiulo2015}; {\sc ySAM}: \citealp{Lee2013}; {\sc Galform}: \citealp{GonzalezPerez2014}; {\sc SantaCruz}: \citealp{Porter2014}; {\sc Lgalaxies}: \citealp{Henriques2015}; {\sc Sage}: \citealp{Croton2016}; {\sc SHARK}: \citealp{Lagos2018}). 

The problem is that, while the ELN criterion determines not very accurately \citep{Athanassoula2008} but reasonably well
whether a disc is stable or not, it does not by itself predict  the mass of the bar or pseudobulge that forms in an unstable disc.
Additional assumptions are needed.

Some SAMs make the extreme assumption that discs evolve into bulges when the ELN criterion is satisfied (e.g. \citealp{GonzalezPerez2014,Gargiulo2015}). Others adopt a more conservative approach and transfer  
to the bulge or pseudobulge just enough matter to make the disc stable again (e.g. \citealp{Henriques2015, Cattaneo2017}). 
Here, we adopt a new approach based on computer simulations.

\citet{Devergne2020} studied the growth of pseudobulges in isolated thin exponential stellar discs embedded in static spherical haloes. 
The novelty of their work was the size of the explored parameter space.
They found that the final bulge-to-total mass ratio $B/T$ could be fitted to an accuracy of $30$ per cent by the formula:
\begin{equation}
\frac{B}{T} = 0.5 f_{\rm d}^{1.8}(3.2r_{\rm d}),
\label{e:BT}
\end{equation}
where $f_{\rm d} = v_{\rm d}^2/v_{\rm c}^2$ is the contribution of the disc to the total gravitational acceleration ($v_{\rm c}$ is the circular velocity at the optical radius $r=3.2r_{\rm d}$, where $r_{\rm d}$ is the exponential scale-length;
$v_{\rm d}$ is the circular velocity considering only the disc's gravity).

We use Eq.~(\ref{e:BT}) to compute the mass $M_{\rm bar}=(B/T)M_{\rm disc}$ of the bar developed by a thin exponential disc of mass $M_{\rm disc}$
when it is embedded in a spherical mass distribution such that $v_{\rm c}$ is the total circular velocity at $r=3.2r_{\rm d}$. The gravitational potential of the spherical mass distribution is
the sum of the contributions from the DM halo, the bulge and the central cusp. If there is no bar, a bar of mass $M_{\rm bar}$ is created from  stars and gas from the disc.
Stars and gas are transferred from the disc to the bar in a proportion that reflects their contributions to the total disc mass.

The case with a pre-existing bar is more complicated because our guidance, Eq.~(\ref{e:BT}), is based on simulations that did not contemplate that possibility.
We compute $v_{\rm d}(3.2r_{\rm d})$ as if the bar were part of the disc, which is not unreasonable, since the formation of a bar should have no substantial impact on the rotation curve at the optical
radius. If the $M_{\rm bar}$ computed with Eq.~(\ref{e:BT}) is larger than the mass $M_{\rm bar,0}$ of the pre-existing bar,
we transfer a mass $M_{\rm bar}-M_{\rm bar,0}$ from the disc to the bulge.

\subsection{Mergers}
\label{Mergers}

Decades of computer simulations have shown that nearly-equal-mass (major) mergers result in the formation of elliptical galaxies and drive gas into the central region, where it fuels starbursts and can feed the growth of supermassive BHs
\citep{Toomre1972,Barnes1988,Barnes1996,
Hopkins2006}. The structure of merger remnants depends not only on the mass ratio of the merging galaxies but also on the initial conditions of the encounters
\citep{Negroponte1983,Martin2018} and the gas content of the merging galaxies (discs can survive even equal-mass mergers if the gas fraction is large enough; \citealp{Hopkins2009}).
Minor mergers, in which a small satellite plunges into a much larger galaxy, heat the disc of the larger galaxy and can form a thick-disc component \citep{Statler1988, Quinn1993}. However, the heating efficiency scales as the square of the mass ratio.
Hence, a Milky Way galaxy may experience $\sim 5$ to 10 mergers with a mass ratio of 1:10 and still retain a spiral morphology \citep{Hopkins2008b}.

The challenge for SAMs is to find a simple analytic model that captures the essence of these complex dynamical transformations.
Here, we consider two models for the morphologies of the merger remnants.
The first is the same as in {\sc GalICS}~2.0 \citep{Cattaneo2017} and {\sc GalICS}~2.1 \citep{Cattaneo2020}. It is the standard assumption in SAMs of galaxy formation.
The second is a novelty of {\sc GalICS}~2.2.
\\
\\
\textbf{Model $1$}\\
\\
In the so-called {\it threshold} model, a sharp threshold $\mu_{\rm m}$  in the mass ratio  $\mu = {\cal{M}}_2/ {\cal{M}}_1$ of the merging galaxies
(with ${\cal{M}}_2<{\cal{M}}_1$) distinguishes major and minor mergers. 
Bulges form in major mergers only and major mergers destroy discs completely.
Variations of the threshold model remain the standard description of mergers in most SAMs to date ({\sc Galacticus}: \citealp{Benson2012}; {\sc Morgana}: \citealp{LoFaro2009}; {\sc Sag}: \citealp{Gargiulo2015}; {\sc ySAM}: \citealp{Lee2013}; {\sc Galform}: \citealp{GonzalezPerez2014}; {\sc Lgalaxies}: \citealp{Henriques2015}; {\sc Sage}: \citealp{Croton2016}; {\sc SHARK}: \citealp{Lagos2018}). We now explain our implementation.

In our SAM, ${\cal{M}}_1$ and ${\cal{M}}_2$ are the total (baryonic plus DM) masses within the baryonic half-mass radii of the merging galaxies.
In major mergers ($\mu>\mu_{\rm m}$), all the stars form a large bulge; all the cold galactic gas is transferred to the central cusp, where it fuels a starburst. 
In minor mergers ($\mu<\mu_{\rm m}$), the gas and the stars in the disc, the bar, the bulge and the cusp of the smaller (secondary) galaxy are added to the disc, the bar, the bulge and the cusp of the larger (primary) galaxy, respectively, without any changes
in the structural properties of the latter (the sizes and characteristic speeds for the four components remain the same). We assume a threshold ratio of $\mu_{\rm m}=0.25$ as in \citet{Cattaneo2017,Cattaneo2020}.
\\
\\
\textbf{Model $2$}\\
\\
Our second model is based on the results of \citet{Kannan2015}, who used high-resolution hydrodynamical simulations to study the formation of classical bulges and the triggering of starbursts in mergers. 
Let $f_{\rm gas,disc}$ be the gas fraction in the disc of the primary galaxy.
We assume that during a merger:
\begin{enumerate}
\item a fraction $\mu$ of the stars in the disc of the primary galaxy is transferred to the central bulge;
\item another fraction $0.2\mu$ of the same stars is scattered into the  halo, which acquires in this way a
stellar component in our SAM; the same applies to the stars of the secondary galaxy that accrete onto the disc of the merger remnant;
\item a fraction  $\mu (1-f_{\rm gas,disc})$ of the gas in the disc of the primary galaxy is transferred to the central cusp (gas falls to the centre less easily than stars if $f_{\rm gas,disc}$ is significant); 
\item even the gas that remains in the disc undergoes a starburst in the case of a major merger ($\mu>0.25$); we assume that this gas has the same SF timescale as that in the central cusp;
\item the stars and the gas in the bar of the primary galaxy are transferred to the bulge and the cusp of the merger remnant, respectively;
\item a fraction $\mu$ of all the stars of the secondary galaxy ends up in the bulge of the merger remnant;  the rest is added to the disc;
\item all the gas of the secondary galaxy ends up in the central cusp. 
\end{enumerate}

Assumptions (i), (ii), (iii) and (vi) are based on the quantitative results of \citet{Kannan2015}\footnote{\citet{Kannan2015} find  $f_{\rm db}=0.3\mu-0.4\mu$, where $f_{\rm db}$ is the fraction of the primary's stellar disc that ends up in the bulge of the merger remnant. Using high-resolution hydrodynamic simulations, \citet{Hopkins2009} find $f_{\rm db}=\mu$, instead. We adopt the latter value here (assumption i), as we find it to be in better agreement with the observed bulge-to-total stellar mass ratios.}. Assumption (iv) is in agreement with observations \citep{Pan2019} and
simulations \citep{Powel2013,Kannan2015}: most major mergers (including the Antennae; \citealp{Wang2004}) exhibit extended SF in their early stages,
which then concentrates to the nuclear region as the galaxies approach coalescence.
Our assumption that all the gas is exhausted during major mergers, whether it is transferred to the central cusp or remains in the disc, may be too strong. On the other hand,
minor mergers \citet{Kannan2015} and fly-by events (e.g. \citet{Lamastra2013}) can induce starbursts, too, and we do not consider these effects in our SAM.

\citet{Kannan2015} did not consider the presence of bars  and thus could not comment on their evolution (assumption v). \citet{Bournaud2002} and \citet{Berentzen2007}  found that the formation of a bulge and dynamical heating 
can prevent even gas-rich systems from retaining a bar after a major merger. 
The transfer of stars from the bar to the bulge  during mergers is an assumption that has no effect on our results because it changes neither the total stellar masses nor the SFRs of galaxies, but
what happens to the gas in the bar is important.
Our assumption that it moves to the cusp is based on an observationally-motivated consideration:
bars funnel gas into the central
region, where it settles into a nuclear mini-spiral, which the interaction with a companion can easily destabilise (see, e.g., NGC 1097; \citealp{Prieto2019}). The latter assumption is crucial for BH growth and quenching in our model, since our SAM does not currently include a model for BH accretion via disc instabilities.

\citet{Kannan2015} did not comment on the fate of the gas in the secondary galaxy either (assumption vii), but we have tested different models, and we have found,   in agreement with \citet{Hopkins2009}, that 
out treatment of the  gas in the secondary galaxy has no statistically significant effect on the evolution of galaxies.

\subsection{Supermassive black hole growth}
We assume that BHs are fed by  the cold gas that accumulates in the central cusp during mergers. The formation of a BH is triggered when the mass of this gas exceeds our minimum BH mass
$M_{\rm seed}=10^2\,{\rm M}_\odot$ (consistent with stellar mass BHs formed from the collapse of massive Population III stars; e.g. \citealp{Latif2016,Haemmerle2020}). If both  galaxies have pre-existing BHs, 
we assume that they instantaneously coalesce, so that any accretion occurs onto the remnant.
 
The BH growth rate is given by:
\begin{equation}
\dot{M}_{\rm BH} = (1-\epsilon_{\rm rad})\mathrm{min}(
\epsilon_{\rm BH}{\rm SFR}_{\rm cusp}, \dot{M}_{\rm Edd}),
\label{e:BHSFR}
\end{equation}
where $\epsilon_{\rm rad}=0.1$ is the radiative efficiency of BH accretion, $\epsilon_{\rm BH}$ is a parameter of the SAM
that sets the efficiency of BH accretion relative to that of SF in the cusp component, SFR$_{\rm cusp}$ is the SFR in the central cusp,
and
$\dot{M}_{\rm Edd}=M_{\rm BH}{\rm c}^2/t_{\rm S}$ is the Eddington rate, where c is the speed of light and $t_{\rm S}$ is the \citet{Salpeter1964} timescale.

In the previous versions of our SAM, $\epsilon_{\rm BH}$ was a constant (\citealp{Cattaneo2017} used $\epsilon_{\rm BH} =0.01$). However, the observations show that $\dot{M}_{\rm BH}/{\rm SFR}$ depends on
 $M_*$  (e.g. \citealp{ Rodighiero2015,Delvecchio2015}). Hydrodynamic simulations reproduce this behaviour and show that it can be explained as a consequence of supernova feedback.
 Supernovae blow the gas out and thus prevent its accretion onto BHs in systems with low escape speeds
 \citep{Dubois2015,Angles2017,Bower2017,Habouzit2017,Davies2019b,Dekel2019,Oppenheimer2020}.

Here, we follow \citet{Kauffmann2000} and assume that:
\begin{equation}
\epsilon_{\rm BH} = \frac{\epsilon_{\rm BH,max}}{1+(v_{\rm BH}/\upsilon_{\rm vir})^2},
\label{e:BHgrowth}
\end{equation}
where $\epsilon_{\rm BH,max}$ and $v_{\rm BH}$ are free parameters that control the maximum BH accretion efficiency and the virial velocity at which the efficiency saturates, respectively.
This choice prevents the formation of overly massive BHs in low-mass, gas-rich galaxies.

\citet{Kauffmann2000} and \citet{Croton2006} fitted the observations with $\epsilon_{\rm BH,max}=0.03$ and $v_{\rm BH}=280\:{\rm km/s}$.
We adopt the same $v_{\rm BH}$ as theirs, but a higher $\epsilon_{\rm BH,max}=0.05$, since our efficiency is relative to SFR$_{\rm cusp}$ and not to the total SFR\footnote{ In \citet{Kauffmann2000}, all the cold gas present after a merger was availaible for BH growth. In \citet{Croton2006}, the mass of the available gas was reduced by a factor equal to the merger ratio $\mu$, as in our model, but there was no additional factor $1-f_{\rm gas, disc}$ (assumption iii)).
Our BHs have a smaller accretion reservoir. They need to accrete more efficiently to reach the same final masses.}.
The case for $\epsilon_{\rm BH,max}=0.05$ can be justified {\it a posteriori}, with the argument that it fits the $M_{\rm BH}$--$M_*$ relation at $z=0$, but it can also be made {\it a priori}
(directly from observations, without running the SAM). \citet{Harris2016} have shown that bright quasars with bolometric luminosities $\gsim 3\times 10^{13}{\rm\,L}_\odot$ (and thus accretion rates $\gsim 20{\rm\,M}_\odot{\rm\,yr}^{-1}$)
 at $2<z<3$ live in galaxies with SFRs of 400--$500{\rm\,M}_\odot{\rm\,yr}^{-1}$. It is reasonable to presume that, in such systems, most of the SF comes from the starburst mode (e.g. Fig. 3 of \citealp{Cattaneo2013}). That gives a BH accretion rate-to-SFR ratio of $\epsilon_{\rm BH} \sim 20/400=0.05$, in agreement with our best-fit value for $\epsilon_{\rm BH,max}$.

In {\sc GalICS}~2.2, mergers are the only channel for BH growth. 
We do not consider accretion via disc instabilities or direct accretion of gas from the hot halo.
However, the role of disc instabilities is indirectly accounted for through the assumption (v) of model~2 in Section \ref{Mergers}. The accretion of hot gas 
is important to explain the population of weak radio galaxies, i.e. Fanaroff-Riley type I radio sources \citep{fanaroff_riley74}, but it makes a small contribution to BH masses
(e.g. \citealp{cattaneo_best09}). 

\subsection{Quenching}
\label{Quenching}
Observationally, the probability that a central galaxy is passive correlates strongly  with both $M_{\rm vir}$ and $M_{{\rm BH}}$ (e.g. \citealp{Bluck2020}). We thus consider two different models for the quenching of galaxies.\\
\\
\textbf{Model A}\\
\\
The notion of halo quenching has its origin in the work of \citet{Dekel2006} and \citet{Cattaneo2006}.
In our current implementation (model A), cold-mode accretion, hot-mode accretion and reaccretion from the galactic fountain are shut down when $M_{\rm vir}>M_{\rm crit}$, the latter being a free parameter of the SAM.
 \citet{Dekel2006} identified $M_{\rm crit}$ with the critical halo mass above which a stable shock heats the infalling cold intergalactic gas, giving rise to a hot quasi-hydrostatic atmosphere.
 However, shock heating does not necessarily cause quenching if it is not accompanied by a mechanism that prevents the shock-heated gas from cooling down again (e.g. \citealp{Cattaneo2020}).

\citet{Cattaneo2006} proposed that BHs couple to the hot gas and that this coupling provides this mechanism.  As soon as a hot atmosphere appears, the growth of BHs becomes self-regulated.
In the self-regulated regime, BHs accrete slowly and release most of their power mechanically, via radio jets.
The power released is just enough to compensate the radiative losses to X-rays, so that the hot gas is kept in thermodynamic equilibrium (\citealp{Cattaneo2007,Cattaneo2009}, and references therein).

Halo quenching  can thus act as an implicit model for radio-mode AGN feedback.
\citet{Cattaneo2006} reproduced the colour--magnitude distributions of galaxies in the Sloan Digital Sky Survey in quantitative detail
with a shutdown of cooling above $M_{\rm crit}=2\times 10^{12}{\rm\,M}_\odot$.
\\
\\
\textbf{Model B}\\
\\
\citet{Silk1998} proposed a model in which BHs grow until the energy deposited in the surrounding gas is larger than the one required to unbind 
the entire baryonic content of their host haloes. 
Energy injection is not the only mechanism through which AGN can drive winds and blow away the gas in their host galaxies (momentum can drive them, too; \citealp{Fabian1999,King2003}).
Energy-driving is, however, the most common assumption \citep{Ostriker2005, Booth2010, Booth2011, Bower2017, Oppenheimer2018, Davies2019, Davies2020, Oppenheimer2020}
and appears in better agreement with recent observational developments (see below).

\begin{figure}
\includegraphics[width=1.\hsize]{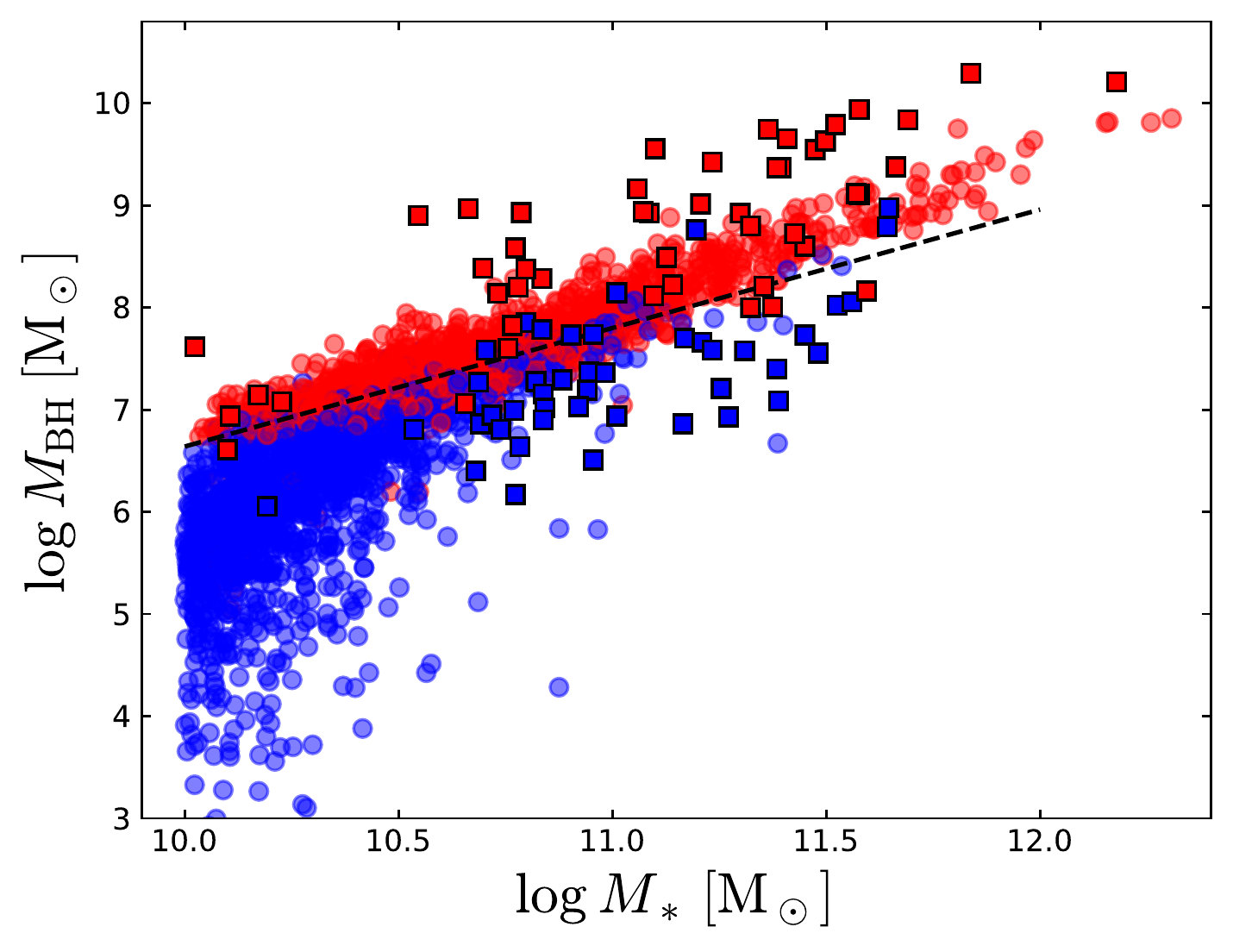} 
\caption{Local BH mass--stellar mass relation for star-forming (${\rm sSFR}>10^{-11}\:{\rm yr^{-1}}$; blue) and passive (${\rm sSFR}<10^{-11}\:{\rm yr^{-1}}$; red) galaxies in model $B$2 (circles) and the observations \citep[squares]{Terrazas2016}. 
The black dashed line shows the quenching boundary between the star-forming and the passive population (Eq.~\ref{e:modelB}).}
\label{i:MsBHSF}
\end{figure}

\citet{Chen2020} derived a similar model, which they called the binding-energy model, from an observational rather than theoretical perspective. Using a simple analytic model, they found that they could reproduce several scaling relations for star-forming and passive galaxies if galaxies cross into the green valley when the time-integrated energy output of their central BHs exceeds a multiple $f_{\rm bind}$ of the gravitational binding energy of the baryons within the halo, that is, when:
\begin{equation}
\label{e:Chen}
\epsilon_{\rm BH\,fb} M_{\rm BH}{\rm c}^2 > f_{\rm bind} E_{\rm bind} = f_{\rm bind} \times \frac{1}{2} f_{\rm b} M_{\rm vir} v_{\rm vir}^2, 
\end{equation}
where  $\epsilon_{\rm BH\,fb}$ is the efficiency for converting BH accretion  into the useful feedback energy and $c$ is the speed of light.
The important point here is that, although $\epsilon_{\rm BH\,fb}$ and $f_{\rm bind}$ are considerably uncertain, only their ratio $\epsilon_{\rm eff}=\epsilon_{\rm BH\,fb}/f_{\rm bind}$ matters to determine the critical BH mass: 
\begin{equation}
M_{\rm BH}^{\rm crit}={f_{\rm b}\over 2\epsilon_{\rm eff}}M_{\rm vir}\bigg({v_{\rm vir}\over c}\bigg)^2,
\label{e:modelB}
\end{equation}
above which a galaxy is quenched.
Eq.~(\ref{e:modelB}) translates into a relation between $M_{\rm BH}^{\rm crit}$ and $M_*$ because of the tight $M_*$--$M_{\rm vir}$ relation at a given $z$.
We treat its normalization $\epsilon_{\rm eff}$ as a free parameter of the SAM and set it to $\epsilon_{\rm eff}=0.00115$ to reproduce the quenching boundary $M_{\rm BH}^{\rm crit}(M_*)$ that separates passive and star-forming galaxies on the $M_{\rm BH}$--$M_*$ diagram at $z=0$ (Fig.~\ref{i:MsBHSF}; the dashed curve has the same normalisation as in Fig.~2c of \citealp{Chen2020}).

\begin{figure*}
\begin{center} 
\includegraphics[width=1\hsize]{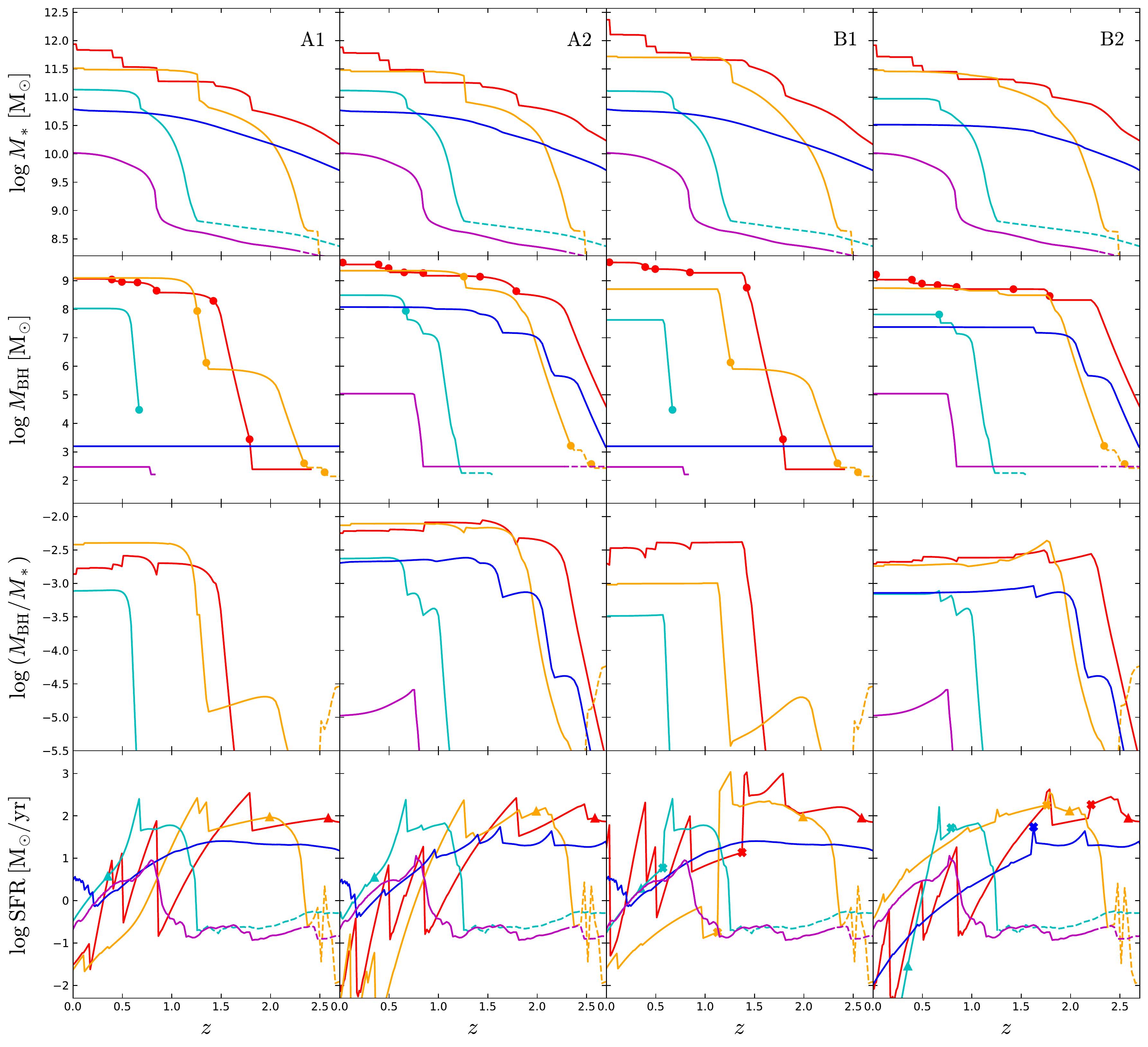}
 \end{center}
\caption{Evolutionary tracks of five galaxies in the four different models considered in this work (from left to right: $A$1, $A$2, $B$1 and $B$2), on the $M_*$--$z$, $M_{\rm BH}$--$z$, $M_{\rm BH}/M_*$--$z$ and SFR--$z$ plane. Dashed lines indicate periods when galaxies are satellites, while solid lines indicate periods when galaxies are centrals (a galaxy that was a satellite at some point in the past and is central at $z=0$ is a backsplash galaxy, i.e., a galaxy that has passed through a group or a cluster and come out to the other side). Points in the second row indicate major mergers, which correspond to the strongest bursts of SF visible in the fourth row. Minor mergers are not noted due to their high frequency. Colored triangles in the fourth row show the point when each galaxy is quenched in the halo quenching model $A$, i.e., when $M_{\rm vir}>2 \times 10^{12}\:{\rm M_\odot}$, while X marks show when a galaxy is quenched in model $B$1 (third panel) and $B$2 (fourth panel), i.e., when it satisfies the BH-quenching criterion~(\ref{e:modelB}).}
\label{i:Evol}
\end{figure*}

As soon as the condition in Eq.~(\ref{e:modelB}) is satisfied, all the gas within the central components of the galaxy (the central cusp, the bulge and the bar,
but not the disc) is reheated and transferred instantaneously to the hot atmosphere. So is the gas within the cold filaments and the galactic fountain, after which
the cooling of hot gas is permanently shut down. The subsequent evolution of the galaxy is characterized by a decreasing SFR as the cold gas left over in the disc is depleted, unless gas-rich mergers reactivate SF and BH growth (Fig.~\ref{i:Evol}).

Our assumption  that the cold gas heated by quasar feedback mixes with the hot atmosphere was made to conserve mass (we had to put it somewhere, be it within the halo or in the intergalactic medium) but has no bearing
on the properties of galaxies, since we do not let it reaccrete.
It would matter, however, if we extended the scope of our analysis from galaxies to the circumgalactic medium by comparing to X-ray observations of the hot gas in early-type galaxies or if we addressed the maintenance problem instead of just assuming that, after the initial quenching, accretion is permanently shut down.

Physically, it is reasonable to suppose that the effects of quasar feedback should depend on the depth of the gravitational potential well.
In a low-mass galaxy, quasar feedback may empty the halo of its entire gaseous content if the condition in Eq.~(\ref{e:modelB}) is satisfied.
In a group, that is a more difficult proposition.
In a cluster, it is downright impossible (we see the gas in X-rays).

Observationally, a recent study \citep{hou_etal21} has analysed archival Chandra X-ray data for 57 field early-type galaxies in the mass range $10^9\,{\rm M_\odot}\lsim M_*\lsim 10^{11}\,{\rm M_\odot}$.
Only in seven of those (all at $M_*>10^{11}\,{\rm M_\odot}$), there is a firm detection of extended X-ray emission and only one (NGC 3193) has an X-ray halo that extends beyond the starlight distribution.
Retrospectively and without consequences for the conclusions of this article, it thus seems that a complete expulsion of all the gas in the halo would have been a more realistic assumption, at least for $M_*< 10^{11}\, {\rm M_\odot}$.

To facilitate a better understanding of the models considered in this work, Fig.~\ref{i:Evol} compares the evolution of five galaxies (two giants, two with masses comparable to that of the Milky Way, and one of low mass)
in models $A$1, $A$2, $B$1, $B$2. For halo quenching ($A$), the merger model has little impact on the evolution of $M_*$.
For BH quenching ($B$), quenched galaxies end up with higher masses in merger model $1$ than in merger model $2$ (in model $B$1, quenching occurs later because one has to wait for a major merger).
The final stellar masses in model $B$2 are similar to those in models $A$1 and $A$2 for giant galaxies but slightly lower at intermediate masses. 

The merger model affects $M_{\rm BH}$ more than the quenching mechanism does. In model $1$, BHs form and grow in major mergers only (circles), although a galaxy can also inherit a BH 
from a smaller companion in a minor merger (magenta curve in models $A$1 and $B$1). In model $2$, minor mergers, too, play a role; BHs grow faster; galaxies are quenched earlier and there is less SF at $z=0$.

In giant galaxies (red and orange curves), halo quenching usually occurs earlier than BH quenching. At intermediate masses (cyan and blue curves), the opposite is true.
The galaxy corresponding to the blue curve is never quenched in model $A$. At lower masses ($M_* \approx 10^{10}\:{\rm M_\odot}$), galaxies are commonly not quenched in any model (see also Fig.~\ref{i:morph}).
Even in that case, however, the SFR can decline at low $z$ if the accretion rate onto the halo declines (magenta curve). In the same way that the SFR can decline without quenching, the opposite is also possible. Quenching does not always mark the end of SF, which can be reactivated by gas-rich mergers.

\subsection{Environmental effects: ram-pressure stripping}

Stripping is the removal of matter from satellite galaxies or their circumgalactic media through environmental effects, that is,
 tides and ram pressure. Tides can take place due to interactions with the gravitational potential of the host halo
(tidal stripping; \citealp{Merritt1984, Byrd1990, Mayer2006}) or to smaller-scale
galaxy-galaxy interactions (harassment; \citealp{Farouki1981, Moore1996, Duc2008}).  
Tidal stripping affects not only  gas but also DM and stars, but ram pressure strips the gas more effectively than tides do (e.g. \citealp{Simpson2018}).

Independently of its physical mechanism (tides or ram pressure), stripping manifests differently depending on its intensity.
Weak stripping deprives satellite galaxies of their circumgalactic media. Without an accretion reservoir, their SFR declines gradually until all the cold gas is exhausted.
Proposed by \citet{Larson1980} to explain the high ratio of S0  to spiral galaxies in clusters, this evolutionary picture has often been referred to as ``strangulation" \citep{Balogh2000} or ``starvation" \citep{Bekki2002}.
Strong stripping removes matter from within galaxies, starting from their outer parts \citep{Gunn1972, Abadi1999}.
The jellyfish galaxies \citep{Ebeling2014} are the most extreme observational evidence of this phenomenon.

In early SAMs, there was no accretion onto satellite galaxies. This assumption is equivalent to an extreme form of strangulation and leads to predicting that nearly all satellites are passive,
at odds with observations \citep{Weinmann2006,Baldry2006,Kang2008,Kimm2009} and cosmological simulations \citep{CattaneoBlaizot2007}.
\citet{Font2008} and \citet{Guo2011,Guo2013} improved the agreement with the observations by considering that 
the circumgalactic media of satellite galaxies are stripped gradually.
\citet{Henriques2015} found that the ram-pressure stripping (RPS) model of \citet{Guo2013} worked well for clusters of galaxies but was too efficient and thus produced too many quiescent satellites in groups.

In the standard versions of {\sc GalICS~2.0} \citep{Cattaneo2017}  and {\sc GalICS~2.1} \citep{Cattaneo2020}, there was no explicit model for
tides or RPS, but tidal stripping was partially accounted for  through its effects on the DM (haloes stop growing and accreting baryons once they become subhaloes; e.g. \citealp{Tollet17}).
In {\sc GalICS~2.0}, this tidally-induced starvation was more than sufficient to shut down gas accretion  onto satellites because galaxies grew through cold accretion only and cold filaments disappear quickly
(on a freefall time) if they are not replenished by accretion from the intergalactic medium.
Too many satellites were passive.
Adding stripping (as we did in one of the versions of  {\sc GalICS~2.0} considered in  \citealp{Koutsouridou2019}) would only make the problem  worse.
In {\sc GalICS}~2.1, satellite galaxies can accrete through cooling. Hence, RPS is important to avoid that
mergers with gas-rich satellites reactivate SF in massive central galaxies.
In contrast, the effects of tidal stripping on the gas content of satellites are negligible compared to those of RPS (e.g. \citealp{Simpson2018}).

Our model for RPS is the same as the one described  in  \citet{Koutsouridou2019} and very similar to that by \citet{Cora2018}.
It is based on the notion that stripping  is a two-stage process. The hot gas is stripped first. The removal of cold galactic gas does not begin until
the hot atmosphere has been stripped down to the size of the galaxy \citep{Bekki2009}.

The RPS of hot gas is computed following \citet{McCarthy2008}, who showed that a simple analytic model  for a spherical gas distribution reproduces  remarkably well the stripping measured in hydrodynamic simulations. 
The RPS radius $R_{\rm s,h}$ for the hot gas
 is determined by requiring that the ram pressure exerted by the hot gas of the host on the satellite  should balance the subhalo's gravitational restoring force:
 \begin{equation}
p_{\rm ram}(r) = \rho_{\rm host}(r) v_{\rm sat}^2(r) = \frac{\pi}{2} \frac{{\rm G} M_{\rm sat}(R_{\rm s,h}) \rho_{\rm sat} (R_{\rm s,h})}{R_{\rm s,h}},
\label{pram}
\end{equation}
Here $r$ is the distance of the satellite from the centre of the host system, 
$\rho_{\rm host}(r)$ is the density of the hot atmosphere of the host at the satellite's location,
$v_{\rm sat}(r)$ is the orbital speed of the satellite in the host's reference frame,
G is the gravitational constant
$\rho_{\rm sat}(R_{\rm s,h})$ is the density of the hot atmosphere of the satellite at the stripping radius, and
$M_{\rm sat}(R_{\rm s,h})$ is the mass of the subhalo within that radius; $\rho_{\rm host}$ and $\rho_{\rm sat}$ are obtained by applying Eq.~(\ref{Faltenbacher}) to the host halo and the subhalo, respectively.

The hot gas retained by the satellite is that within $R_{\rm s,h}$. Its mass is given by the volume integral of $\rho_{\rm sat}$ within a sphere of radius $R_{\rm s,h}$.
If this integral returns a value lower than the mass of the hot gas within the subhalo, the difference is stripped from the subhalo and transferred to the host.
If $R_{\rm s,h}$ is smaller that the optical radius of
the satellite (i.e. the radius that contains 83 per cent of the baryonic mass of the galaxy), we assume
that the whole gaseous halo of the satellite is stripped. In our SAM, this is also a necessary condition for the stripping of cold galactic gas.

Our model for RPS of cold gas in the discs of satellite galaxies is based on \citet{Gunn1972}.
The equation for the stripping radius $R_{\rm s,c}$ of the cold gaseous disc is:
\begin{equation}
\label{Gunn}
p_{\rm ram} = \rho_{\rm h}(r) \upsilon_{\rm sat}^2(r) {\rm cos} \theta = 2 \pi{\rm G} \Sigma_{\rm disc}(R_{\rm s,c})\Sigma_{\rm gas}(R_{\rm s,c}).
\end{equation}
Eq.~(\ref{Gunn}) is analogous to Eq.~(\ref{pram}) with two differences. First, the left-hand side depends on the angle
$\theta$ between the disc's rotation axis and the orbital velocity of the satellite galaxy  (that is, the direction of the hot wind that blows the cold gas away if we examine the problem in the reference frame of the satellite).
Second, the restoring force at the right-hand side is now assumed  to be due to the disc's rather than the halo's gravity.
 $\Sigma_{\rm disc}$ and $\Sigma_{\rm gas}$  are  the total surface density (stars plus gas)
 and the surface density of the gaseous disc, respectively (we assume that they have the same exponential scale-length).
Once we have determined the RPS radius $R_{\rm s,c}$ for the cold mass, the calculation of the cold gas  stripped from the satellite proceeds in the same manner as we did for the hot gas.
We compute the mass of the cold gaseous disc within $R_{\rm s,c}$. If this mass is lower than the gas mass of the disc, the difference is removed from the disc and transferred to the hot atmosphere of the host halo.

Several studies have shown that the stripping radii from  Eq.~(\ref{Gunn}) are similar  to those found in hydrodynamical simulations, albeit somewhat smaller (e.g. \citealp{Roediger2007, Tonnesen2009, Steinhauser2016}).
This small discrepancy is  understandable because the inner halo should make some  contribution to the disc's restoring force and also because of the clumpiness of the cold gas, which makes it more resilient to RPS.

\section{Results}
\label{Results}
\begin{figure*}
\begin{center}
\includegraphics[width=0.9\hsize]{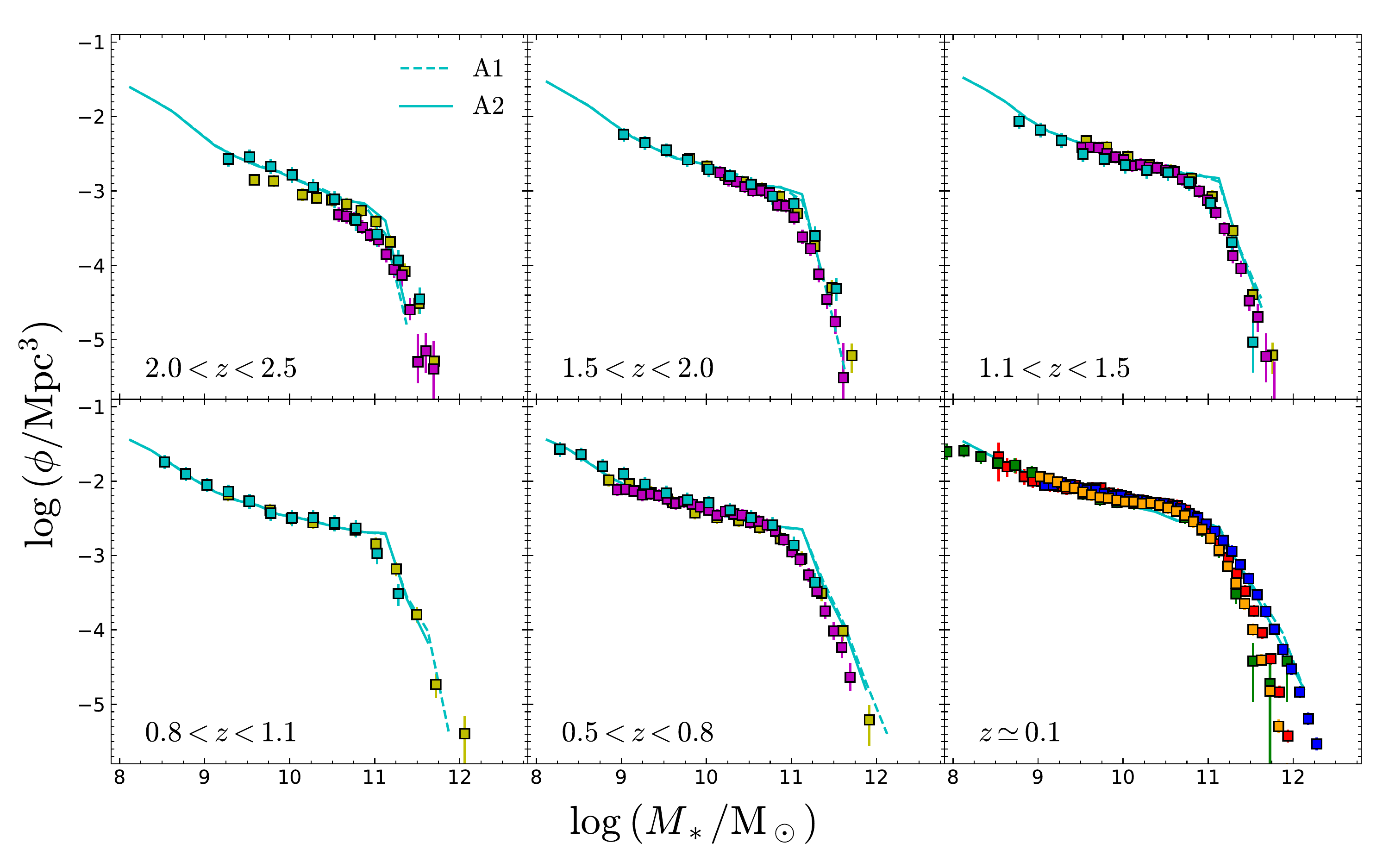}\\
\includegraphics[width=0.9\hsize]{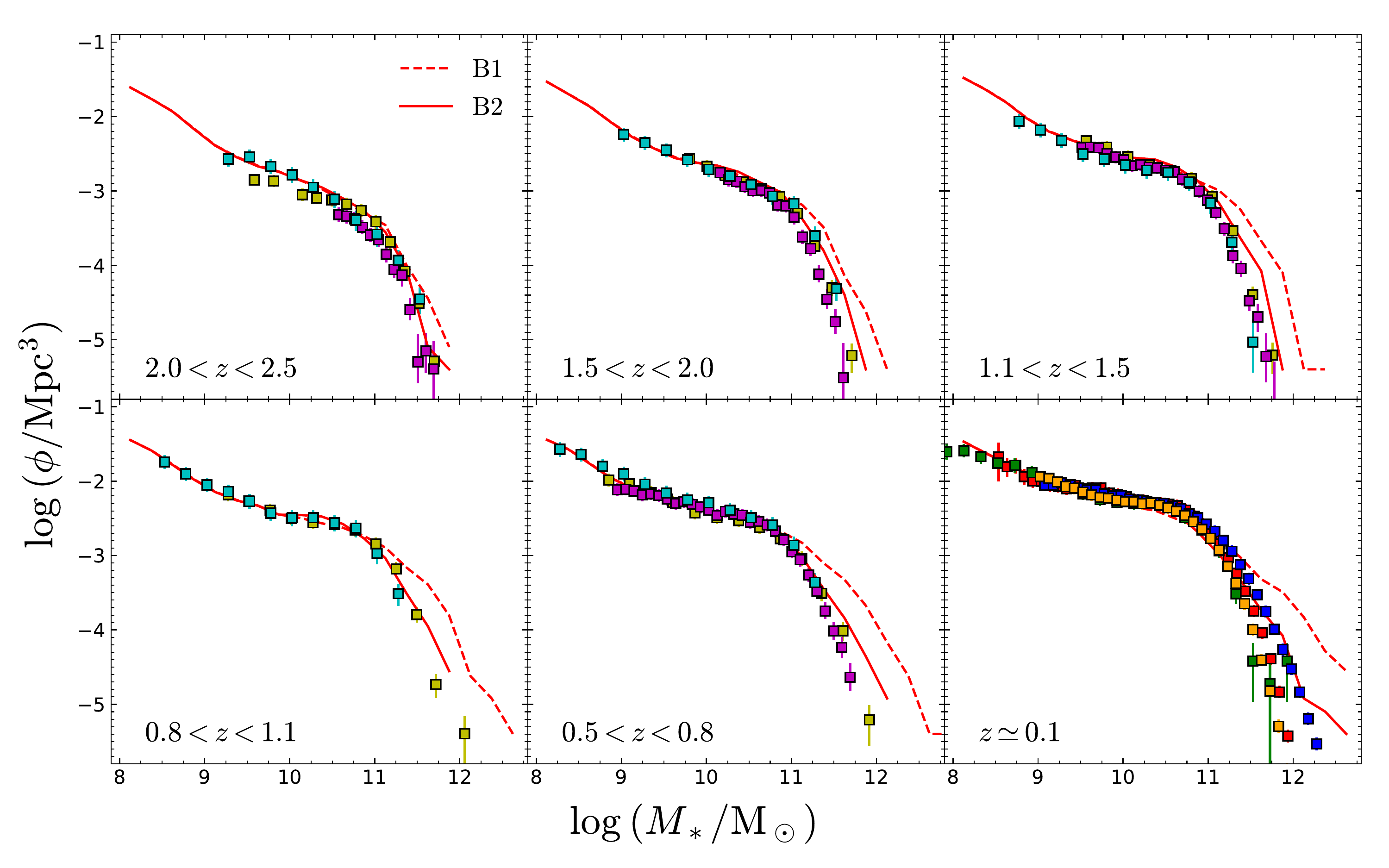} 
\end{center}
\caption{Galaxy stellar mass functions at $0<z<2.5$ for models $A$1 and $A$2 (above;  cyan dotted and solid curves, respectively) and models $B$1 and $B$2 (below; red dotted and solid curves, respectively). The data points show the observations from \citeauthor{Ilbert2013} (\citeyear{Ilbert2013}; yellow), \citeauthor{Muzzin2013} (\citeyear{Muzzin2013}; magenta), \citeauthor{Tomczak2014} (\citeyear{Tomczak2014}; cyan), \citeauthor{Yang2009} (\citeyear{Yang2009}; red), \citeauthor{Baldry2012} (\citeyear{Baldry2012}; green), \citeauthor{Bernardi2013} (\citeyear{Bernardi2013}; blue) and \citeauthor{Moustakas2013} (\citeyear{Moustakas2013}; orange).}
\label{GSMF1}
\end{figure*}

In this section, we compare the predictions of our two quenching models (the halo-quenching model $A$ and the BH/binding-energy-quenching model $B$) when combined with the  merger models 1 and 2 described in Section~\ref{Mergers}.
That makes four models in total: $A$1,  $A$2, $B$1, $B$2.

Fig.~\ref{GSMF1}  focusses on the galaxy stellar mass function (GSMF) at $0<z<2.5$. With halo quenching, the GSMF does not depend on the assumed merger model. Models $A$1 and $A$2 are almost indistinguishable. For $M_{\rm crit}=2\times 10^{12}{\rm\,M}_\odot$,
they are both in good agreement with the observations (data points with error bars)
 at all redshifts. The only discrepancy is at the knee of the GSMFs, where halo quenching predicts a sharp decrease in the number of massive galaxies in contrast to the smoother downturn seen in observations. This feature arises from the abrupt 
 shutdown of all accretion modes  at $M_{\rm crit}$.
 We have run tests and we have verified that we could improve the agreement with the observations by using three different critical masses for the shutdown of cooling, cold accretion and re-accretion from the galactic fountain, but
 here we have tried to keep the model as simple as possible to make our point clearer.
 
With BH/binding-energy quenching, the merger model has a significant impact on the GSMF. Model $B$2 clearly fits the observations much better than model $B$1.
Its only problem is a slight excess in the number densities of very massive galaxies but this could be attributed to the lack of tidal stripping in our SAM
(compare the green curve with tidal stripping and the blue curve without it in \citealp{Tollet17}, Fig.~10, right panel).

\begin{figure*}
\begin{center}
\includegraphics[width=0.9\hsize]{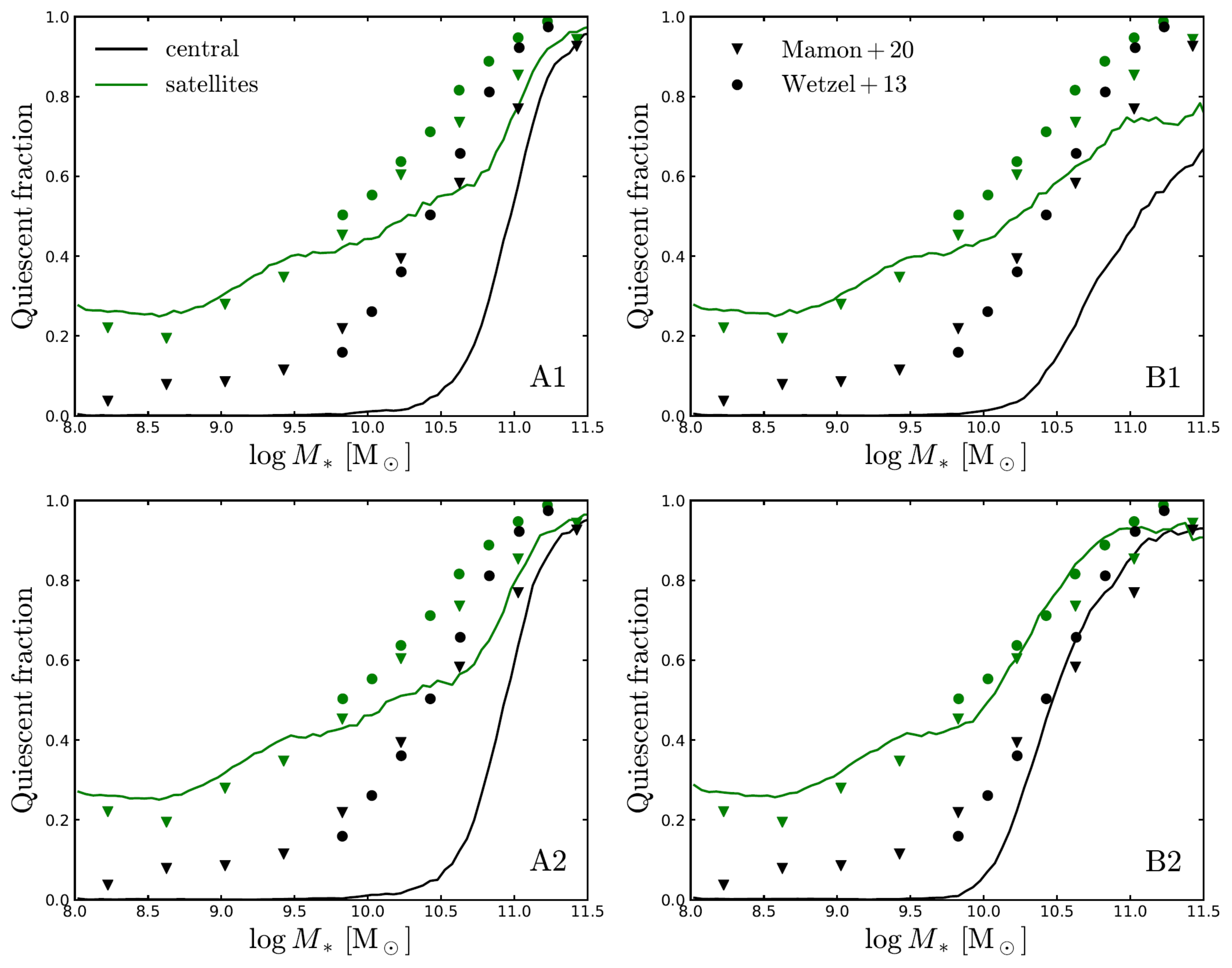}
\end{center}
\caption{
Fraction of passive central (black) and satellite (green) galaxies as a function of stellar mass at $z\sim 0.1$. The left panels show models A1 (top) and A2 (bottom) and right panels show models B1 (top) and B2 (bottom). Data points in all panels show the observations of \citeauthor{Wetzel2013} (\citeyear{Wetzel2013}; circles) and \citeauthor{Mamon2020} (\citeyear{Mamon2020}; triangles). Passive galaxies are defined as those with ${\rm sSFR }={\rm SFR}/M_*<10^{-11} \: {\rm yr}^{-1}$.
}
\label{Qfrac}
\end{figure*}

\begin{figure*}
\begin{center}
\includegraphics[width=0.9\hsize]{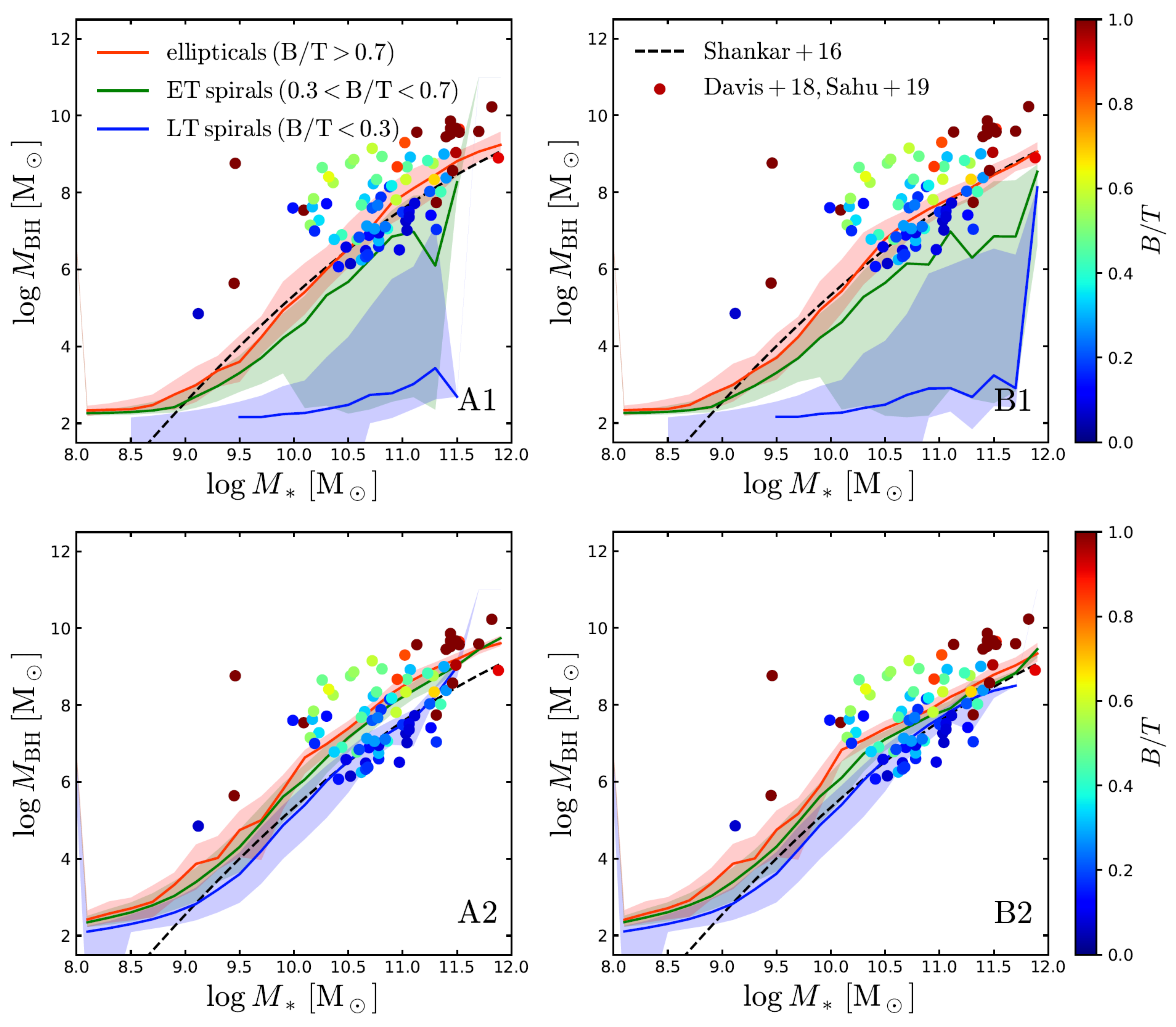} 
\end{center}
\caption{Median central black hole mass as a function of stellar mass, as predicted by the halo quenching models A1 and A2 (top and bottom left panels) and the binding-energy quenching models B1 (top right) and B2 (bottom right panel). The shaded areas indicate the 16th and 84th percentiles of the predicted distributions. Red, green and blue color refer to elliptical galaxies ($B/T>$0.7), early-type spirals (0.3$<B/T<$0.7) and late-type spirals ($B/T<$0.3), respectively. Observations of individual galaxies from \citet{Davis2018} and \citet{Sahu2019} are shown with points, color-coded according to the galaxies' $B/T$. The black, dashed line shows the $M_{\rm BH}$-$M_*$ relation for early-type galaxies (ellipticals and lenticulars) corrected for selection biases from \citet{Shankar2016}.}
\label{BH}
\end{figure*}

The fraction of passive galaxies as a function of $M_*$ (Fig.~\ref{Qfrac}) displays a similar behaviour.
Models $A$1 and $A$2 are, once again, almost indistinguishable. They both fail to reproduce the observed passive fractions of central galaxies at all but the highest masses, that is, at ${\rm log} (M_*/{\rm M_\odot}) \lesssim 11.5$. 
Model $B$1 is even more catastrophic.
Model $B$2 is the only one in reasonable agreement with the passive fractions of both central and satellite galaxies, if we exclude its difficulty at explaining the existence of a small population of passive central dwarf galaxies.
Notice that all models reproduce the fraction of passive satellites below $\sim 10^{10}{\rm\,M}_\odot$, stressing the environmental origin of their quenching mechanisms (i.e. strangulation/RPS).
Fig.~\ref{BH} (BH masses) and Figs.~\ref{i:BT}--\ref{i:Morph} (morphologies) explain why model $B$2 works much better than model $B$1.

\begin{figure}
\includegraphics[width=1.\hsize]{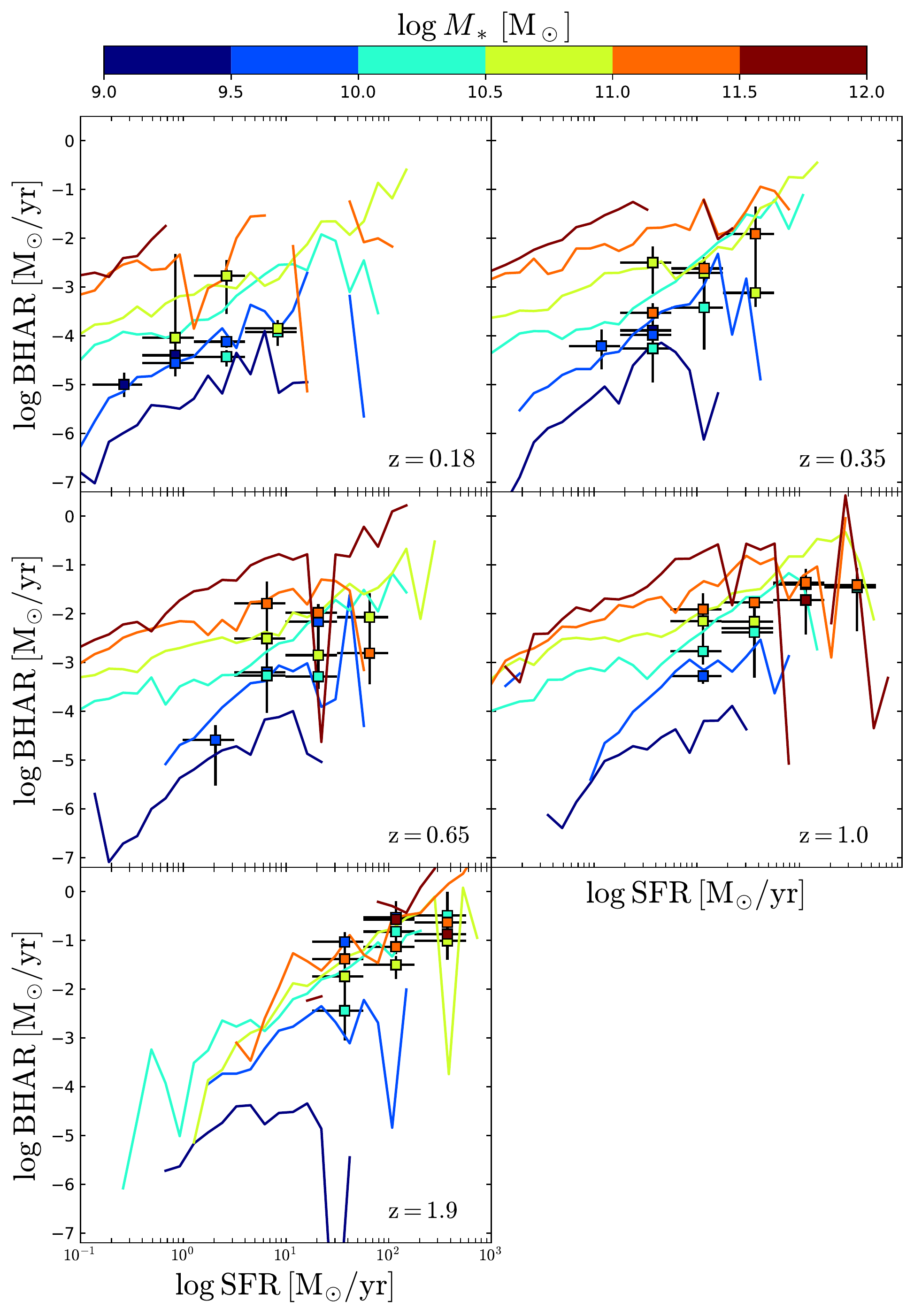} 
\caption{Mean BH accretion rate as a function of the total SFR in {\sc GalICS~2.2}, model $B$2. The colored curves correspond to six bins of stellar mass. The panels to five redshifts.
The points with error bars are the observations  from \citet{Delvecchio2015}, who selected galaxies in the far infrared and measured BH accretion rates from stacked X-ray data. }
\label{i:BHAR}
\end{figure}

 The red, green and blue curves in Fig.~\ref{BH} show the median $M_{\rm BH}$--$M_*$ relation for galaxies of different bulge-to-total stellar mass ratio 
($B/T>0.7$, $0.3<B/T<0.7$ and $B/T<0.3$, respectively).
In all models, $M_{\rm BH}/M_*$ grows with $B/T$ but the trend is far more pronounced in  models $A$1 and $B$1 than it is in models $A$2 and $B$2. The reason is  as follows.
Merger model~1 assumes that, in minor mergers, the gas in the disc and the bar of the secondary galaxy is moved to the disc and the bar of the merger remnant, respectively, and there it stays unless the minor merger is followed by a major one.
So does the gas in the disc and the bar of the primary galaxy. In merger model~2, even minor mergers can transfer gas to the central cusp (Section~2.6) and contribute to the growth of the central BH.
Hence, the quenching condition in Eq.~(\ref{e:modelB}) is more easily satisfied.

Furthermore, if spheroids form through major mergers only, as model~1 assumes,
then repeated minor mergers create a route for the formation of giant spirals.
The fraction of massive galaxies ($M_*\sim 10^{11}{\rm\,M}_\odot$) with $B/T<0.1$ is almost 20 per cent in model $A$1 and more than 20 per cent in model $B$1, but only a few per cent in models $A$2 and $B$2,
in much better agreement with the observations \citep[Fig.~\ref{i:Morph}]{Mendel2014}.
Model $A$1 has fewer massive galaxies with $B/T<0.1$ than model $B$1 because halo quenching prevents merger remnants from reaccreting a disc at $M_{\rm vir}>M_{\rm crit}$.
In model $B$1, it is $M_{\rm BH}$ that counts. Hence, the low 
BH-to-stellar mass ratios of giant spirals result in too much SF at high masses
(Fig.~\ref{Qfrac}) and too many massive galaxies in general  (Fig.~\ref{GSMF1}).

Having analysed and understood how the four models behave, we now we wish to look at the comparison with the observational data in closer detail and especially at the BH masses, which are crucial for model $B$. 
The observational data \citep{Davis2018,Sahu2019}
show that, for a same stellar mass, galaxies with higher $B/T$ host  BHs that are systematically more massive (Fig.~\ref{BH}).
This finding is in qualitative agreement with our theoretical predictions.

The quantitative difference between the $M_{\rm BH}$ of different morphological types in the observations is smaller than our predictions for models $A$1 and $B$1 but larger than our predictions for models $A$2 and $B$2.
This finding suggests that the role of minor mergers and disc instabilities in the growth of supermassive BHs is underestimated in merger model~1 and overestimated in merger model~2.
{\sc GalICS~2.2} does not contain any explicit model for the growth of BHs via disc instabilities but disc instabilities form bars.
The assumption (v) of merger model 2 that the gas in the bar of the primary galaxy falls to the center during mergers is the back door through which disc instabilities affect the growth of BHs in our SAM.
Assuming that all this gas falls to the center even in very minor mergers is too extreme.
Replacing this assumption with a more realistic model for BH growth via disc instabilities is likely to widen the gap between the $M_{\rm BH}$--$M_*$ relations for spiral and elliptical galaxies and thus improve the agreement with the data points.

Despite the above remark, models $A$2 and $B$2 fit the data  much better than models $A$1 and $B$1, especially for late-type galaxies.
In merger model 2,
the $M_{\rm BH}$--$M_*$ relation for galaxies with $B/T<0.3$ overlaps with the observations of \citet[blue circles]{Davis2018}, while merger model 1 is completely off.

For $B/T>0.3$ (red and green curves), model $A$2 fits the data points \citep[circles in the color range from red to green]{Sahu2019} better than model $B$2 because, in model $A$2, BHs keep growing until SF depletes the accretion reservoir
even if the quenching criterion is satisfied.
In model $B$2, the gas in the central cusp is instantaneously ejected once the quenching criterion is satisfied (some BH accretion is nevertheless possible after quenching if mergers with gas-rich satellites replenish the cusp; Fig.~\ref{i:Evol}).

We could bring model $B$2 in better agreement with the BH masses of early-type galaxies by increasing the normalisation $\epsilon_{\rm BH, max}$ of the efficiency of BH accretion while lowering the efficiency $\epsilon_{\rm eff}$ of BH feedback,
so that the fraction of passive galaxies and the GSMF stay the same.
However, that would have two drawbacks. First, the  dashed line in Fig.~\ref{i:MsBHSF} would no longer be at the boundary that separates the red and the blue squares.
Second, the change would bring model $B$2 in conflict  with direct measurements of the BH accretion rate-to-SFR ratio. 

\citet{Delvecchio2015} have studied $\sim 8600$ star-forming galaxies selected from surveys in the far-infrared with Herschel and measured their BH accretion rates from stacked X-ray data (Chandra).
The data points with error bars in Fig.~\ref{i:BHAR} show their findings.  The BH accretion rate-to-SFR ratio increases with $M_*$. {\sc GalICS}~2.2 (model $B$2, colored curves) reproduces this trend.
In both the model and the observations, the highest stellar masses drop out of the lowest redshift panels because the galaxies with the highest masses are the first to become passive (downsizing).
Quantitatively, the BH accretion rates for galaxies with $10^{10.5}{\rm \, M_\odot}<M_*<10^{11}{\rm\,M}_\odot$ are in reasonable agreement with the observations at all $z$.
At $10^{11}{\rm\,M}_\odot<M_*<10^{11.5}{\rm\,M}_\odot$, the model and the observations are in reasonably good agreement at $z\ge 0.65$. At $z\le 0.35$, however, BH accretion rates are {\it overpredicted} both at $M_*>10^{11}{\rm\,M}_\odot$
(red, orange curves) and at $10^{10}{\rm\,M}_\odot<M_*<10^{10.5}{\rm\,M}_\odot$ (cyan curves). 
That is the opposite to what one expects if the efficiency of BH accretion is underestimated.

A bias in the observed BH masses may explain the tension between Fig.~\ref{i:MsBHSF}  and Fig.~\ref{i:BHAR}.
Dynamical mass estimates, such as those used by  \citet{Sahu2019},  require that  the gravitational sphere of influence of the BH should be resolved.
\citet{Shankar2016} argued that  this requirement introduces a bias in favour of galaxies with higher BH masses. Model $B$2 is in reasonably good agreement with  \citet{Shankar2016}'s de-biassed  $M_{\rm BH}$--$M_*$ relation
(Fig.~\ref{BH}, black dashed curve). Alternatively, the mean BH accretion rate may be underestimated because of the relatively small size of \citet{Delvecchio2015}'s sample compared to the rarity of the quasar phenomenon
(in the local Universe, there is one quasar every $\sim 10^4$ early-type galaxies; \citealp{wisotzki_etal01}).

\begin{figure*}
\begin{center}
\includegraphics[width=0.9\hsize]{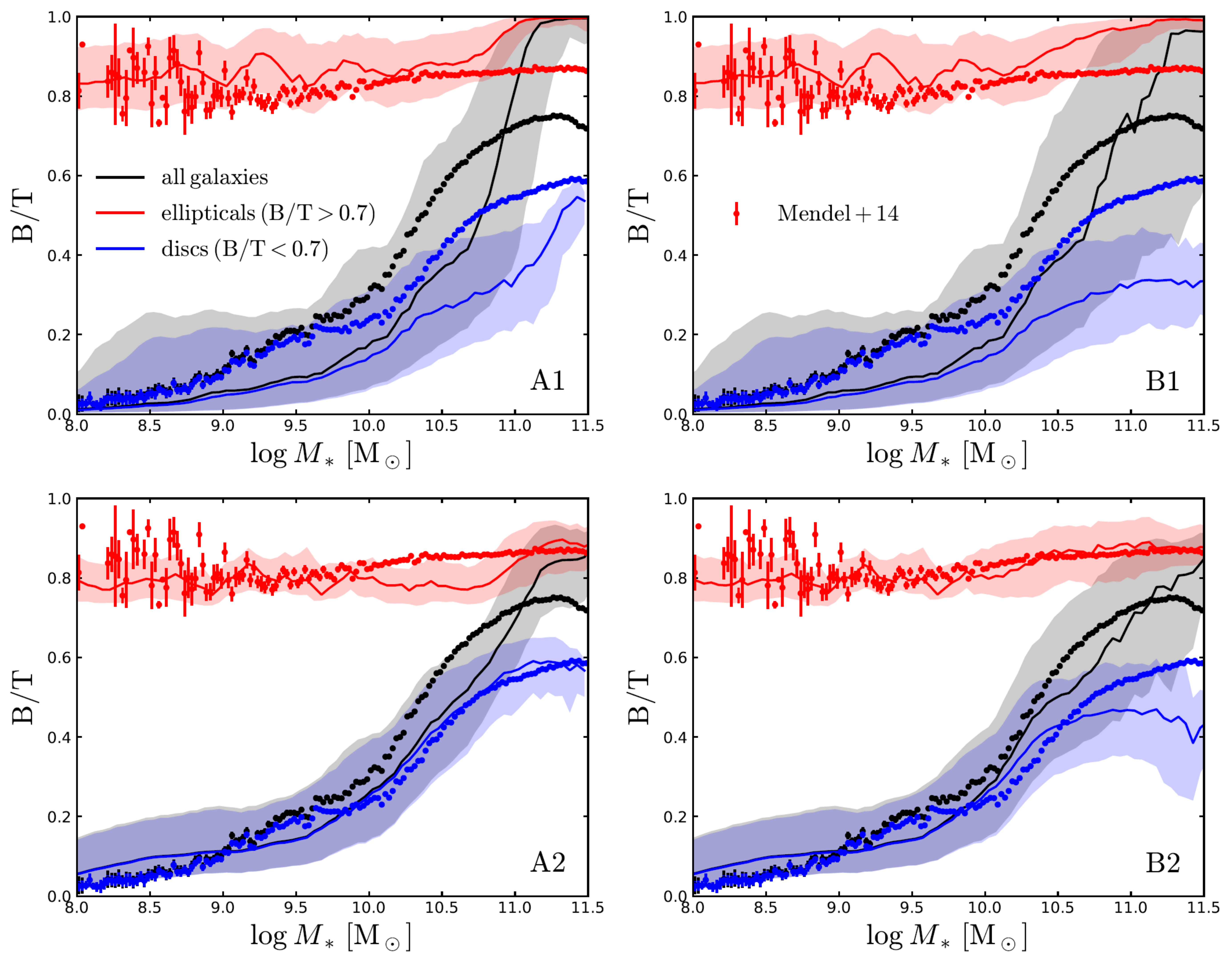} 
\end{center}
\caption{Median bulge-to-total stellar mass ratio as a function of galaxy stellar mass for the two halo quenching models (left panels) and the two binding-energy quenching models (right panels), at $z~0.1$. The shaded areas indicate the 25th and 75th percentiles
of the predicted distribution. Points with errorbars show the observations by \citet{Mendel2014}. Black, red and blue color refer to all galaxies in the sample, ellipticals ($B/T > 0.7$) and disc galaxies ($B/T < 0.7$), respectively.}
\label{i:BT}
\end{figure*}

\begin{figure*}
\begin{center}
\includegraphics[width=0.9\hsize]{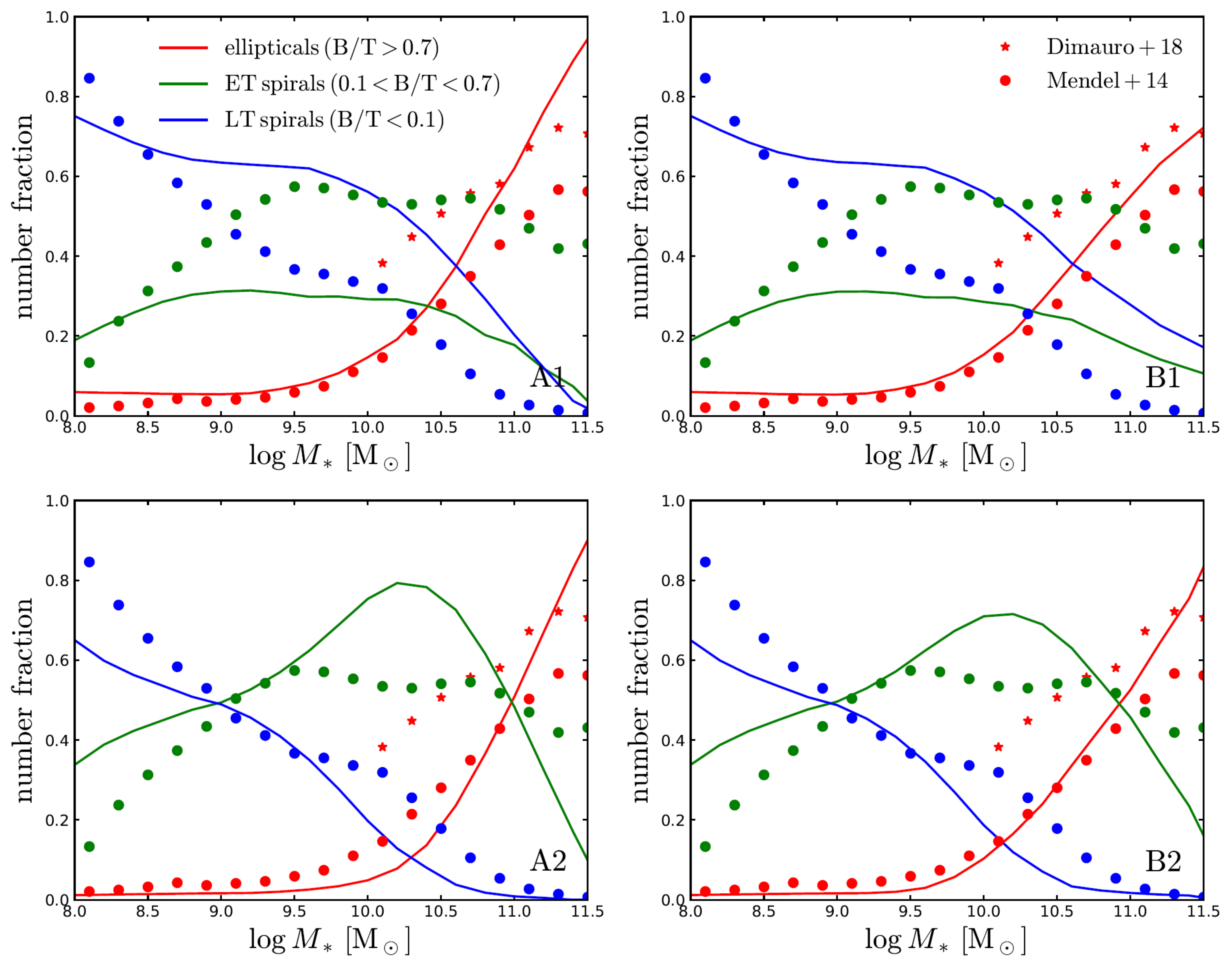} 
\end{center}
\caption{Fraction of elliptical galaxies ($B/T>0.7$; red lines), earlier-type spirals ($0,1<B/T<0.7$; green lines) and late-type spirals ($B/T<0.1$; blue lines) as a function of stellar mass, for the two halo quenching models (left panels) and the two binding-energy quenching models (right panels), at $z\sim 0.1$. The colored points show the observations by \citeauthor{Mendel2014} (\citeyear{Mendel2014}; circles) and \citeauthor{Dimauro2018} (\citeyear{Dimauro2018}; stars).}
\label{i:Morph}
\end{figure*}

Another factor than can alter $M_{\rm BH}$ and thus the final galaxy stellar masses is the assumed initial BH mass $M_{\rm BH,seed}$ (models in which AGN quenching depends on $\dot{M}_{\rm BH}$ rather than
$M_{\rm BH}$ are less sensitive on $M_{\rm BH,seed}$; e.g. \citealp{Somerville2008}).
BHs with $M_{\rm BH,seed} \simeq 100\: {\rm M_\odot}$ are expected to have formed at high redshift (z $\sim 15 - 30$) from the collapse of massive Population III stars (stellar mass BHs; \citealp{Latif2016, Haemmerle2020}).
Observations of extremely bright quasars at $z\sim 6 - 7.5$ hosting supermassive BHs with $M_{\rm BH}\sim 10^{8}$--$10^{10}\:{\rm M_\odot}$ (e.g. \citealp{Fan2006, Wu2015, Banados2018}) are, however, in tension with this BH formation scenario if subsequent BH growth is dominated by Eddington-limited accretion. Models aiming at resolving this issue invoke super-Eddington accretion  or the formation of massive BH seeds ($\sim 10^{5}-10^{6}\:{\rm M_\odot}$) either via direct collapse of metal-poor gas in protogalaxies \citep{Inayoshi2019, Woods2019} or during major mergers of disc-dominated galaxies \citep{Mayer2010}.

Adopting a higher $M_{\rm BH,seed}$  for all galaxies in our SAM results in a dearth of galaxies with 
$M_*\lesssim 10^{10}\:{\rm M_\odot}$. However, if massive seed BHs dominate mostly the massive end of the mass function (as proposed by e.g. \citealp{Bonoli2014}), then our adopted BH/binding-energy quenching model would be in even better agreement with the observed GSMF, as that would solve the problem at the high-mass end.

Galactic morphologies are important both as a test of the merger model (1 vs. 2) and because of the $M_*$--morphology relation.
In Fig.~\ref{i:Morph}, we compare the results of our four models to the observations of \citeauthor{Mendel2014} (\citeyear{Mendel2014}; circles) and \citeauthor{Dimauro2018} (\citeyear{Dimauro2018}; stars) for the morphological breakdown of the galaxy population. In this figure, but not in the rest of the article, we classify galaxies using $B/T$ ratios that consider only the classical bulge. The reason is that \citet{Mendel2013} fit bulges with a \citet{deVaucouleurs1948} model, which is a poor fit to pseudobulges, and assign $B/T = 0$ to discs hosting pseudobulges where ’no de Vaucouleurs component is a fit’.
Once again, the agreement is best for merger model 2. Since merger model 1 does not allow for any bulge growth during minor mergers, models $A$1 and $B$1 predict too many late-type spirals with $B/T<0.1$
(Fig.~\ref{i:Morph}) and average $B/T$ ratios that are too low (Fig.~\ref{i:BT}, blue curves). 

In contrast to halo quenching, BH quenching also affects galaxies at intermediate masses ($10^{10}\: {\rm M}_\odot \lesssim M_* \lesssim 10^{11}\: {\rm M}_\odot$) by preventing  gas accretion onto the disc component.
This is why, in this mass range, the median $B/T$ is higher in model $B$2 than in model $A$2 (Fig.~\ref{i:BT}, black curves).
In Fig.~\ref{i:Morph}, the fraction of galaxies with $0.1<B/T<0.7$ (green curve) is lower in model $B$2 than in model $A$2 at intermediate masses, while the fraction of galaxies with $B/T>0.7$ (red curve) is larger, 
because BH-quenching has prevented merger remnants from regrowing a disc. In both cases, model $B$2 is in better agreement with the observations than model $A$2.
These considerations do not apply when comparing models $A$1 and $B$1 because in model $B$1 there are not enough passive galaxies at $M_* <10^{11}\: {\rm M}_\odot$ for them to affect the
statistical properties of the galaxy population  (Fig.~\ref{Qfrac}).

\begin{figure*}
\begin{center}
\includegraphics[width=0.9\hsize]{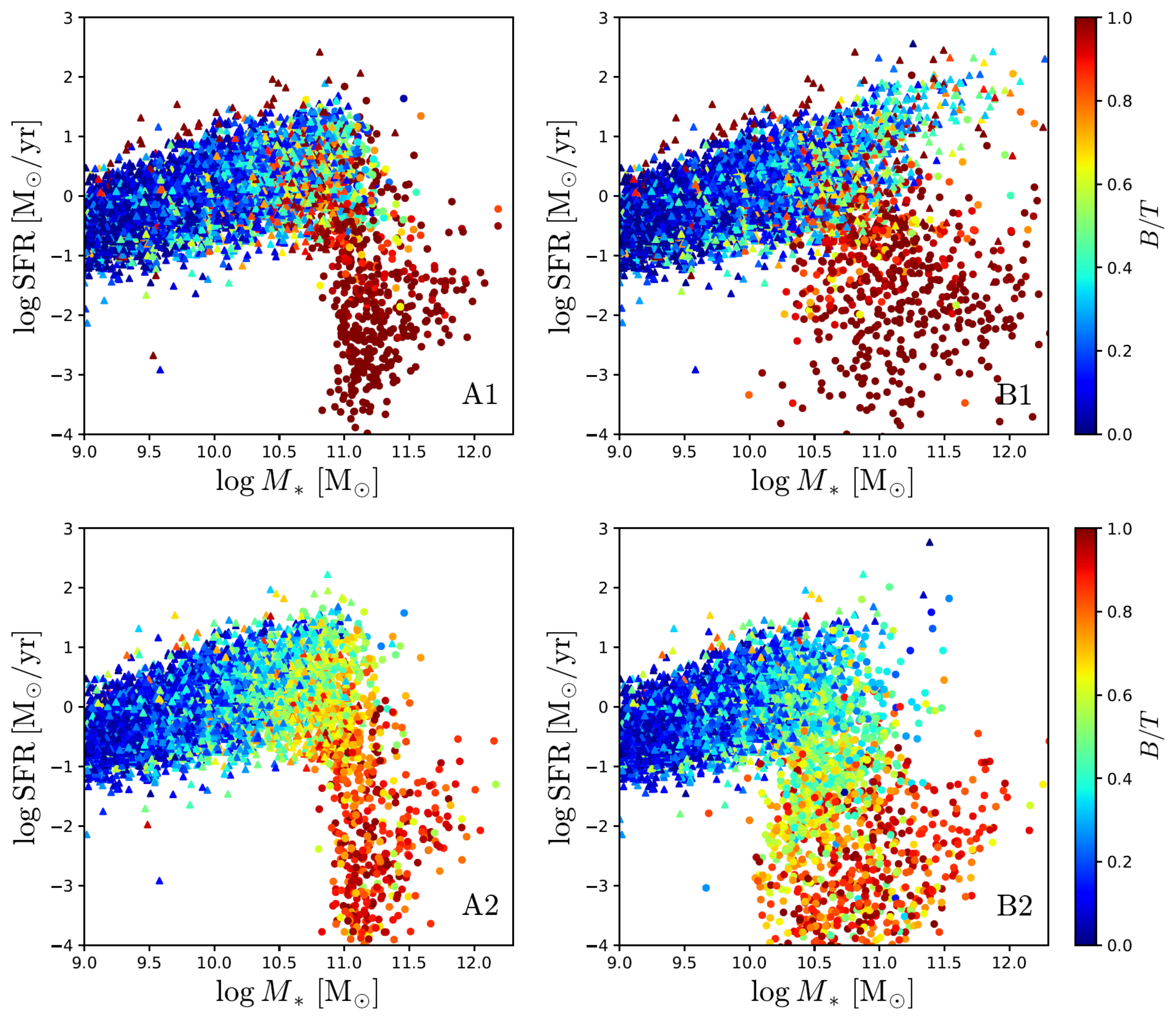} 
\end{center}
\caption{Star formation rate as a function of stellar mass for individual central galaxies in the four models considered, color-coded according to the galaxies' bulge to total mass ratios $B/T$, at $z \sim 0.1$. Circles denote galaxies that have already been quenched, i.e. have satisfied the $M_{\rm vir}>2 \times 10^{12}\:{\rm M_\odot}$ criterion for models $A1$ and $A2$ and the binding-energy criterion (Eq.~\ref{e:modelB}) for models $B1$ and $B2$. Triangles denote galaxies that have not been quenched yet.}
\label{i:morph}
\end{figure*}

Fig.~\ref{i:morph} uses  the morphological distribution of galaxies on the SFR--$M_*$ plane to compare the evolution in the halo and BH quenching scenarios. Triangles  and circles represent unquenched and quenched galaxies, respectively. Quenched galaxies are those that satisfy or have satisfied our quenching criterion, which is  $M_{\rm vir}>2 \times 10^{12}\:{\rm M}_\odot$   for models $A$1 and $A$2 and Eq.~(\ref{e:modelB}) for models $B$1 and $B$2. 
We note that a galaxy may be quenched but still star-forming if it has not consumed all its gas yet (there are circles on the main sequence of star-forming galaxies).
It is also possible that an unquenched galaxy may be passive (there are a few triangles below the main sequence of star-forming galaxies).
This rare phenomenon occurs when a galaxy ceases to accrete and runs out of gas because its halo has stopped growing.

Models $A$1 and $B$1 are once again the most catastrophic, showing too few late-type spirals and a red sequence populated almost completely by galaxies with $B/T\sim 1$.
Furthermore, in model $B$1 the main sequence of star-forming galaxies extends too far at high masses because, in merger model~1, there are too many massive spirals
with BH masses that are not large enough to induce quenching.

In model $A$2,  disc instabilities and mergers cause $B/T$ to increase gradually towards high-masses along  the main sequence of star-forming galaxies  until galaxies reach the stellar mass-scale at which $M_{\rm vir}=M_{\rm crit}=2\times 10^{12}{\,\rm M}_\odot$. Above $M_*=10^{11}\:{\rm M}_\odot$, nearly all galaxies  are quenched and most of them are passive (there is a time lag between quenching and the actual end of SF).
In model $B$2, $B/T$ increases not only towards higher masses on the main sequence but also towards lower SFRs at fixed $M_*$. 
The green valley is not only a region of intermediate SFRs but also a region of intermediate morphologies.
Model $B$2 is in better agreement with the observations (e.g. \citealp{Wuyts2011, Whitaker2015, Morselli2017}; Dimauro et al. 2022).

There are two reasons why model $A$2 fails on this point and model $B$2 is successful.
First, to have a transition from spiral to elliptical galaxies along a vertical line of fixed $M_*$, there has to be a spiral population in the main sequence of star-forming galaxies at that $M_*$.
In model $A$2, quenching starts at $M_*\sim 10^{11}\:{\rm M}_\odot$, where the main sequence is dominated by lenticular ($B/T\sim 0.5$) rather than spiral morphologies.
In model $B$2, galaxies with $M_*>3\times 10^{10}{\rm\,M}_\odot$ often contain BHs that are massive enough for the BH-quenching criterion~(\ref{e:modelB}) to be satisfied. 
Hence, the green valley starts at lower masses, where there is a significant spiral population.
Lowering $M_{\rm vir}^{\rm crit}$ in model $A$2 would result in a more extended green valley, in better agreement with the observations, but would also give too few massive galaxies at all redshifts.
However, the extent of the green valley is not the only problem and here comes the second point.
In model $A$2 there is no causal connection between quenching and morphology. The only connection is that they are both related to mass. Hence, no correlation should be expected between SFR and $B/T$ at constant $M_*$ beyond the one introduced by mergers, which may considerably accelerate the depletion of gas in an already quenched galaxy (this mechanism explains
why, at a fixed $M_*$, the blue circles  are statistically above the red circles in all models).
In model $B$2, quenching is linked to BH mass.  For a same $M_*$, galaxies with larger $B/T$ have higher BH masses (Fig.~\ref{BH}).
Therefore, it is natural that higher $B/T$ should carry a higher quenching probability.

\section{Discussion and Conclusion}

We have investigated the mechanisms that quench star formation in galaxies by comparing
the results of two competing scenarios: $A$) halo quenching and $B$) BH/binding-energy quenching. In model $A$, quenching occurs when $M_{\rm vir}$ grows above the critical halo mass $M_{\rm vir}^{\rm crit}$, a free parameter whose order of magnitude is determined by \citet{Dekel2006}’s critical mass for shock-heating. In model $B$, quenching occurs when $M_{\rm BH}$ grows above the critical BH mass $M_{\rm BH}^{\rm crit}(v_{\rm vir})$ determined by \citet{Chen2020}’s binding-energy criterion.

Model $B$ is more complicated than model $A$ because BH masses, unlike halo masses, depend on the assumptions of the galaxy-evolution model, notably those for  morphological evolution, since the growth of supermassive BHs is related to the formation of bulges. Some of these assumptions are degenerate: a same galaxy can have the same BH mass with a lower bulge-to-total stellar mass ratio and a higher BH-to-bulge mass ratio. Hydrodynamic simulations can assist us to make reasonable assumptions, but, even if one doubted their guidance, there are enough observational constraints to break the aforementioned degeneracy.

We have considered two versions of models $A$ and $B$: one in which only {\it major} mergers lead to the growth of BHs and bulges (models $A$1 and $B$1), the other in which the effects of major and minor mergers are modelled following the numerical results of \citet[models $A$2 and $B$2]{Kannan2015}. Model $B$2 is in better agreement with the observations than the other three. For BH accretion rates equal to $\sim 5$ per cent of the SFR in the central starburst, model B2 reproduces the observed BH masses, galaxy stellar mass functions, quiescent fractions and morphological properties of galaxies.

Three conclusions follow from these findings:
\begin{enumerate}
\item{In agreement with \citet{Terrazas2016} and  \citet{Bluck2020}, {\it quenching is more closely related to the growth of} $M_{\rm BH}$ than it is to that of $M_{\rm vir}$. 
Halo quenching fails to produce enough passive galaxies at $M_* < 10^{11.5}{\rm\,M}_\odot$ (Fig.~\ref{Qfrac}, panels $A$1 and $A$2).
Lowering $M_{\rm vir}^{\rm crit}$ alleviates the problem but destroys the agreement with the galaxy stellar mass function at high masses.
BH-quenching model $B$2 does not present this problem because, 
even at $M_* < 10^{11.5}{\rm\,M}_\odot$, there are some galaxies with $M_{\rm BH}>M_{\rm BH}^{\rm crit}$. 
Halo quenching also fails to reproduce the SFR--morphology relation. With halo quenching, the transition to higher $B/T$ occurs along the main sequence of star-forming galaxies; the green valley is composed of early-type galaxies. With BH-quenching, the transition on the SFR-$M_*$ diagram is much more vertical; at constant
$M_*$, $B/T$ increases in the direction of lower SFRs.}
\item{Morphological evolution leads to quenching because  mergers transform discs into bulges while feeding the growth of supermassive BHs. 
{\it Restricting the formation of bulges and supermassive BHs to major mergers is incompatible with the fraction of passive galaxies at high masses}. There are always galaxies that reach high masses through a series of minor mergers. If these galaxies are not quenched, there are too many star-forming galaxies at high masses.}
\item{If the gas fraction fed to the central BHs in mergers is the same for massive and low-mass systems, then
we have too many passive galaxies at low masses. 
 {\it The efficiency with which gas is fed to the central BH must be lower at lower} $v_{\rm vir}$. This finding is consistent with the proposal that supernova feedback prevents the efficient growth of supermassive BHs in systems with low escape speeds \citep{Dubois2015,Angles2017,Bower2017,
Habouzit2017,Davies2019b,Dekel2019,
Oppenheimer2020}.}
\end{enumerate}

We end with a couple of comments on how these findings fit into our broader understanding of BH feedback.
The first is that our BH quenching criterion (model $B$) is a criterion for $M_{\rm BH}$, while previous SAMs \citep{Croton2006,Somerville2008,Henriques2015} had assumed that  $\dot{M}_{\rm BH}$ is the relevant quantity to determine whether gas accretion is shut down or not.
This assumption is largely based on the insight from cool-core clusters.
In these systems, the intracluster medium has a short cooling time but is not able to cool and flow efficiently onto the central galaxy because  mechanical heating by radio jets from active galactic nuclei compensates the radiative losses to X-rays.
\citep{Peterson2006,Mcnamara2007,Mcnamara2012,
Cattaneo2009, HlavecekLarrondo2022}.
The requirement that the jet power should compensate the X-ray luminosity $L_X$ translates into a criterion for the BH accretion rate required to shut down cooling: $\epsilon_{\rm jet}\dot{M}_{\rm BH}c^2\sim L_X$, where $\epsilon_{\rm jet}$ is the efficiency with which the rest-mass energy of the gas that accretes onto the BH is converted into jet kinetic energy.

As a good theory should explain as many observations as possible with the fewest possible assumptions, it is logical that SAMs should try and extend the lesson from  cool-core clusters to lower masses.
In this case, however, Occam's razor does not cut it. \citet{Chen2020}  and we have shown that we need a different criterion to explain the quenching of SF in galaxies. 
We have not modelled the processes through which BHs quench SF. Hence, we cannot comment on the physics of BH feedback. 
We remark, however, that the criterion required for the initial quenching of SF is different from the one required to maintain it shut down afterwards.
We can, therefore, make the argument that quenching and maintenance correspond to different feedback regimes.
In fact, a  criterion for ${M}_{\rm BH}$ arises naturally in models where BHs accrete at the Eddington limit, so that the accretion power grows with ${M}_{\rm BH}$
until it is large enough to blow away all the gas in the host system \citep{Silk1998,Fabian1999,King2003}.

The second comment is on the entropy excesses on group scales (the X-ray-emitting gas has higher entropies in groups than in clusters when all entropies are rescaled to the virial one;  \citealp{edge_stewart91,evrard_henry91,kaiser91,ponman_etal96,valageas_silk99,lloyd_etal00}).
Already \citet{evrard_henry91} made an explicit argument that these excesses have no reason to be present if shock heating from gravitational collapse is the dominant mechanism driving the evolution of the hot gas, while quasar feedback provides a possible explanation for their existence.
Hence, their observation disfavours model $A$, in which gravitational shock heating quenches SF. It also disfavours any model in which the only BH feedback is maintenance feedback: if BHs provided just enough heat to maintain the X-ray-emitting gas in thermal equilibrium, then BHs would prevent the hot gas from dropping below the entropy at which it was shock-heated but would not raise its entropy above it either.
The entropy excesses on group scales are evidence of a past epoch during which the gas was heated violently. Violent heating implies rapid BH growth. Hence, it is reasonable to associate it with the quasar-mode feedback that quenched SF.

In conclusion, our findings fit naturally in a scenario in which BH feedback is bimodal.  In star-forming galaxies, BHs grow rapidly (quasar mode) until they are massive enough to blow away all the gas  (most early-type galaxies with $10^9{\rm\,M}_\odot\lsim M_*\lsim 10^{11}{\rm\,M}_\odot$ lack extended X-ray emission; \citealp{hou_etal21}) or heat it to high entropy.
In this second regime, the accretion is self-regulated. BHs grow just enough to maintain the hot gas in thermal equilibrium.

\section*{Acknowledgements}

The authors acknowledge interesting conversation with Paola Dimauro and Gary Mamon. IK acknowledges support from the ERC Starting Grant NEFERTITI H2020/808240.

\section*{DATA AVAILABILITY}

The data underlying this article will be shared on reasonable request to the corresponding author.

\bibliographystyle{mn2e}

\bibliography{ref_av}

\begin{thebibliography}{}

\bibitem[\protect\citeauthoryear{{Abadi}, {Moore} \& {Bower}}{{Abadi}
  et~al.}{1999}]{Abadi1999}
{Abadi} M.~G.,  {Moore} B.,    {Bower} R.~G.,  1999, \mnras, 308, 947

\bibitem[\protect\citeauthoryear{{Angl{\'e}s-Alc{\'a}zar},
  {Faucher-Gigu{\`e}re}, {Quataert}, {Hopkins}, {Feldmann}, {Torrey}, {Wetzel}
  \& {Kere{\v{s}}}}{{Angl{\'e}s-Alc{\'a}zar} et~al.}{2017}]{Angles2017}
{Angl{\'e}s-Alc{\'a}zar} D.,  {Faucher-Gigu{\`e}re} C.-A.,  {Quataert} E.,
  {Hopkins} P.~F.,  {Feldmann} R.,  {Torrey} P.,  {Wetzel} A.,    {Kere{\v{s}}}
  D.,  2017, \mnras, 472, L109

\bibitem[\protect\citeauthoryear{{Athanassoula}}{{Athanassoula}}{2008}]{Athanassoula2008}
{Athanassoula} E.,  2008, \mnras, 390, L69

\bibitem[\protect\citeauthoryear{{Ba{\~n}ados}, {Venemans}, {Mazzucchelli},
  {Farina}, {Walter}, {Wang}, {Decarli}, {Stern}, {Fan}, {Davies}, {Hennawi},
  {Simcoe}, {Turner}, {Rix}, {Yang}, {Kelson}, {Rudie} \&
  {Winters}}{{Ba{\~n}ados} et~al.}{2018}]{Banados2018}
{Ba{\~n}ados} E.,  {Venemans} B.~P.,  {Mazzucchelli} C.,  {Farina} E.~P.,
  {Walter} F.,  {Wang} F.,  {Decarli} R.,  {Stern} D.,  {Fan} X.,  {Davies}
  F.~B.,  {Hennawi} J.~F.,  {Simcoe} R.~A.,  {Turner} M.~L.,  {Rix} H.-W.,
  {Yang} J.,  {Kelson} D.~D.,  {Rudie} G.~C.,    {Winters} J.~M.,  2018, \nat,
  553, 473

\bibitem[\protect\citeauthoryear{{Baldry}, {Balogh}, {Bower}, {Glazebrook},
  {Nichol}, {Bamford} \& {Budavari}}{{Baldry} et~al.}{2006}]{Baldry2006}
{Baldry} I.~K.,  {Balogh} M.~L.,  {Bower} R.~G.,  {Glazebrook} K.,  {Nichol}
  R.~C.,  {Bamford} S.~P.,    {Budavari} T.,  2006, \mnras, 373, 469

\bibitem[\protect\citeauthoryear{{Baldry}, {Driver}, {Loveday}, {Taylor},
  {Kelvin}, {Liske}, {Norberg}, {Robotham}, {Brough}, {Hopkins}, {Bamford},
  {Peacock}, {Bland-Hawthorn} \& et al.}{{Baldry} et~al.}{2012}]{Baldry2012}
{Baldry} I.~K.,  {Driver} S.~P.,  {Loveday} J.,  {Taylor} E.~N.,  {Kelvin}
  L.~S.,  {Liske} J.,  {Norberg} P.,  {Robotham} A.~S.~G.,  {Brough} S.,
  {Hopkins} A.~M.,  {Bamford} S.~P.,  {Peacock} J.~A.,  {Bland-Hawthorn} J.,
  et al. 2012, \mnras, 421, 621

\bibitem[\protect\citeauthoryear{{Baldry}, {Glazebrook}, {Brinkmann},
  {Ivezi{\'c}}, {Lupton}, {Nichol} \& {Szalay}}{{Baldry}
  et~al.}{2004}]{Baldry2004}
{Baldry} I.~K.,  {Glazebrook} K.,  {Brinkmann} J.,  {Ivezi{\'c}} {\v{Z}}.,
  {Lupton} R.~H.,  {Nichol} R.~C.,    {Szalay} A.~S.,  2004, \apj, 600, 681

\bibitem[\protect\citeauthoryear{{Balogh}, {Navarro} \& {Morris}}{{Balogh}
  et~al.}{2000}]{Balogh2000}
{Balogh} M.~L.,  {Navarro} J.~F.,    {Morris} S.~L.,  2000, \apj, 540, 113

\bibitem[\protect\citeauthoryear{{Barnes}}{{Barnes}}{1988}]{Barnes1988}
{Barnes} J.~E.,  1988, \apj, 331, 699

\bibitem[\protect\citeauthoryear{{Barnes} \& {Hernquist}}{{Barnes} \&
  {Hernquist}}{1996}]{Barnes1996}
{Barnes} J.~E.,  {Hernquist} L.,  1996, \apj, 471, 115

\bibitem[\protect\citeauthoryear{{Barro}, {Faber}, {Koo}, {Dekel}, {Fang},
  {Trump}, {P{\'e}rez-Gonz{\'a}lez}, {Pacifici}, {Primack}, {Somerville},
  {Yan}, {Guo}, {Liu}, {Ceverino}, {Kocevski} \& {McGrath}}{{Barro}
  et~al.}{2017}]{Barro2017}
{Barro} G.,  {Faber} S.~M.,  {Koo} D.~C.,  {Dekel} A.,  {Fang} J.~J.,  {Trump}
  J.~R.,  {P{\'e}rez-Gonz{\'a}lez} P.~G.,  {Pacifici} C.,  {Primack} J.~R.,
  {Somerville} R.~S.,  {Yan} H.,  {Guo} Y.,  {Liu} F.,  {Ceverino} D.,
  {Kocevski} D.~D.,    {McGrath} E.,  2017, \apj, 840, 47

\bibitem[\protect\citeauthoryear{{Bekki}}{{Bekki}}{2009}]{Bekki2009}
{Bekki} K.,  2009, \mnras, 399, 2221

\bibitem[\protect\citeauthoryear{{Bekki}, {Couch} \& {Shioya}}{{Bekki}
  et~al.}{2002}]{Bekki2002}
{Bekki} K.,  {Couch} W.~J.,    {Shioya} Y.,  2002, \apj, 577, 651

\bibitem[\protect\citeauthoryear{{Benson}}{{Benson}}{2012}]{Benson2012}
{Benson} A.~J.,  2012, \na, 17, 175

\bibitem[\protect\citeauthoryear{{Berentzen}, {Shlosman}, {Martinez-Valpuesta}
  \& {Heller}}{{Berentzen} et~al.}{2007}]{Berentzen2007}
{Berentzen} I.,  {Shlosman} I.,  {Martinez-Valpuesta} I.,    {Heller} C.~H.,
  2007, \apj, 666, 189

\bibitem[\protect\citeauthoryear{{Bernardi}, {Meert}, {Sheth}, {Vikram},
  {Huertas-Company}, {Mei} \& {Shankar}}{{Bernardi}
  et~al.}{2013}]{Bernardi2013}
{Bernardi} M.,  {Meert} A.,  {Sheth} R.~K.,  {Vikram} V.,  {Huertas-Company}
  M.,  {Mei} S.,    {Shankar} F.,  2013, \mnras, 436, 697

\bibitem[\protect\citeauthoryear{{Bigiel}, {Leroy}, {Walter}, {Brinks}, {de
  Blok}, {Madore} \& {Thornley}}{{Bigiel} et~al.}{2008}]{Bigiel2008}
{Bigiel} F.,  {Leroy} A.,  {Walter} F.,  {Brinks} E.,  {de Blok} W.~J.~G.,
  {Madore} B.,    {Thornley} M.~D.,  2008, \aj, 136, 2846

\bibitem[\protect\citeauthoryear{{Bigiel}, {Leroy}, {Walter}, {Brinks}, {de
  Blok}, {Kramer}, {Rix}, {Schruba}, {Schuster}, {Usero} \&
  {Wiesemeyer}}{{Bigiel} et~al.}{2011}]{Bigiel2011}
{Bigiel} F.,  {Leroy} A.~K.,  {Walter} F.,  {Brinks} E.,  {de Blok} W.~J.~G.,
  {Kramer} C.,  {Rix} H.~W.,  {Schruba} A.,  {Schuster} K.~F.,  {Usero} A.,
  {Wiesemeyer} H.~W.,  2011, \apjl, 730, L13

\bibitem[\protect\citeauthoryear{{Bluck}, {Maiolino}, {S{\'a}nchez}, {Ellison},
  {Thorp}, {Piotrowska}, {Teimoorinia} \& {Bundy}}{{Bluck}
  et~al.}{2020}]{Bluck2020}
{Bluck} A. F.~L.,  {Maiolino} R.,  {S{\'a}nchez} S.~F.,  {Ellison} S.~L.,
  {Thorp} M.~D.,  {Piotrowska} J.~M.,  {Teimoorinia} H.,    {Bundy} K.~A.,
  2020, \mnras, 492, 96

\bibitem[\protect\citeauthoryear{{Bonoli}, {Mayer} \& {Callegari}}{{Bonoli}
  et~al.}{2014}]{Bonoli2014}
{Bonoli} S.,  {Mayer} L.,    {Callegari} S.,  2014, \mnras, 437, 1576

\bibitem[\protect\citeauthoryear{{Booth} \& {Schaye}}{{Booth} \&
  {Schaye}}{2010}]{Booth2010}
{Booth} C.~M.,  {Schaye} J.,  2010, \mnras, 405, L1

\bibitem[\protect\citeauthoryear{{Booth} \& {Schaye}}{{Booth} \&
  {Schaye}}{2011}]{Booth2011}
{Booth} C.~M.,  {Schaye} J.,  2011, \mnras, 413, 1158

\bibitem[\protect\citeauthoryear{{Bournaud} \& {Combes}}{{Bournaud} \&
  {Combes}}{2002}]{Bournaud2002}
{Bournaud} F.,  {Combes} F.,  2002, \aap, 392, 83

\bibitem[\protect\citeauthoryear{{Bower}, {Benson}, {Malbon}, {Helly}, {Frenk},
  {Baugh}, {Cole} \& {Lacey}}{{Bower} et~al.}{2006}]{Bower2006}
{Bower} R.~G.,  {Benson} A.~J.,  {Malbon} R.,  {Helly} J.~C.,  {Frenk} C.~S.,
  {Baugh} C.~M.,  {Cole} S.,    {Lacey} C.~G.,  2006, \mnras, 370, 645

\bibitem[\protect\citeauthoryear{{Bower}, {Schaye}, {Frenk}, {Theuns},
  {Schaller}, {Crain} \& {McAlpine}}{{Bower} et~al.}{2017}]{Bower2017}
{Bower} R.~G.,  {Schaye} J.,  {Frenk} C.~S.,  {Theuns} T.,  {Schaller} M.,
  {Crain} R.~A.,    {McAlpine} S.,  2017, \mnras, 465, 32

\bibitem[\protect\citeauthoryear{{Brook}, {Stinson}, {Gibson}, {Wadsley} \&
  {Quinn}}{{Brook} et~al.}{2012}]{Brook2012}
{Brook} C.~B.,  {Stinson} G.,  {Gibson} B.~K.,  {Wadsley} J.,    {Quinn} T.,
  2012, \mnras, 424, 1275

\bibitem[\protect\citeauthoryear{{Byrd} \& {Valtonen}}{{Byrd} \&
  {Valtonen}}{1990}]{Byrd1990}
{Byrd} G.,  {Valtonen} M.,  1990, \apj, 350, 89

\bibitem[\protect\citeauthoryear{{Cattaneo} \& {Best}}{{Cattaneo} \&
  {Best}}{2009}]{cattaneo_best09}
{Cattaneo} A.,  {Best} P.~N.,  2009, \mnras, 395, 518

\bibitem[\protect\citeauthoryear{{Cattaneo}, {Blaizot}, {Devriendt} \&
  {Guiderdoni}}{{Cattaneo} et~al.}{2005}]{Cattaneo2005}
{Cattaneo} A.,  {Blaizot} J.,  {Devriendt} J.,    {Guiderdoni} B.,  2005,
  \mnras, 364, 407

\bibitem[\protect\citeauthoryear{{Cattaneo}, {Blaizot}, {Devriendt}, {Mamon},
  {Tollet}, {Dekel}, {Guiderdoni}, {Kucukbas} \& {Thob}}{{Cattaneo}
  et~al.}{2017}]{Cattaneo2017}
{Cattaneo} A.,  {Blaizot} J.,  {Devriendt} J.~E.~G.,  {Mamon} G.~A.,  {Tollet}
  E.,  {Dekel} A.,  {Guiderdoni} B.,  {Kucukbas} M.,    {Thob} A.~C.~R.,  2017,
  \mnras, 471, 1401

\bibitem[\protect\citeauthoryear{{Cattaneo}, {Blaizot}, {Weinberg},
  {Kere{\v{s}}}, {Colombi}, {Dav{\'e}}, {Devriendt}, {Guiderdoni} \&
  {Katz}}{{Cattaneo} et~al.}{2007}]{CattaneoBlaizot2007}
{Cattaneo} A.,  {Blaizot} J.,  {Weinberg} D.~H.,  {Kere{\v{s}}} D.,  {Colombi}
  S.,  {Dav{\'e}} R.,  {Devriendt} J.,  {Guiderdoni} B.,    {Katz} N.,  2007,
  \mnras, 377, 63

\bibitem[\protect\citeauthoryear{{Cattaneo}, {Dekel}, {Devriendt}, {Guiderdoni}
  \& {Blaizot}}{{Cattaneo} et~al.}{2006}]{Cattaneo2006}
{Cattaneo} A.,  {Dekel} A.,  {Devriendt} J.,  {Guiderdoni} B.,    {Blaizot} J.,
   2006, \mnras, 370, 1651

\bibitem[\protect\citeauthoryear{{Cattaneo}, {Faber}, {Binney}, {Dekel},
  {Kormendy}, {Mushotzky}, {Babul}, {Best}, {Br{\"u}ggen}, {Fabian}, {Frenk},
  {Khalatyan}, {Netzer}, {Mahdavi}, {Silk}, {Steinmetz} \&
  {Wisotzki}}{{Cattaneo} et~al.}{2009}]{Cattaneo2009}
{Cattaneo} A.,  {Faber} S.~M.,  {Binney} J.,  {Dekel} A.,  {Kormendy} J.,
  {Mushotzky} R.,  {Babul} A.,  {Best} P.~N.,  {Br{\"u}ggen} M.,  {Fabian}
  A.~C.,  {Frenk} C.~S.,  {Khalatyan} A.,  {Netzer} H.,  {Mahdavi} A.,  {Silk}
  J.,  {Steinmetz} M.,    {Wisotzki} L.,  2009, \nat, 460, 213

\bibitem[\protect\citeauthoryear{{Cattaneo}, {Koutsouridou}, {Tollet},
  {Devriendt} \& {Dubois}}{{Cattaneo} et~al.}{2020}]{Cattaneo2020}
{Cattaneo} A.,  {Koutsouridou} I.,  {Tollet} E.,  {Devriendt} J.,    {Dubois}
  Y.,  2020, \mnras, 497, 279

\bibitem[\protect\citeauthoryear{{Cattaneo} \& {Teyssier}}{{Cattaneo} \&
  {Teyssier}}{2007}]{Cattaneo2007}
{Cattaneo} A.,  {Teyssier} R.,  2007, \mnras, 376, 1547

\bibitem[\protect\citeauthoryear{{Cattaneo}, {Woo}, {Dekel} \&
  {Faber}}{{Cattaneo} et~al.}{2013}]{Cattaneo2013}
{Cattaneo} A.,  {Woo} J.,  {Dekel} A.,    {Faber} S.~M.,  2013, \mnras, 430,
  686

\bibitem[\protect\citeauthoryear{{Cavagnolo}, {Donahue}, {Voit} \&
  {Sun}}{{Cavagnolo} et~al.}{2009}]{Cavagnolo2009}
{Cavagnolo} K.~W.,  {Donahue} M.,  {Voit} G.~M.,    {Sun} M.,  2009, \apjs,
  182, 12

\bibitem[\protect\citeauthoryear{Chabrier}{Chabrier}{2003}]{Chabrier2003}
Chabrier G.,  2003, Publications of the Astronomical Society of the Pacific,
  115, 763

\bibitem[\protect\citeauthoryear{{Chen}, {Faber}, {Koo}, {Somerville},
  {Primack}, {Dekel}, {Rodr{\'\i}guez-Puebla}, {Guo}, {Barro}, {Kocevski}, {van
  der Wel}, {Woo}, {Bell}, {Fang} \& et al.}{{Chen} et~al.}{2020}]{Chen2020}
{Chen} Z.,  {Faber} S.~M.,  {Koo} D.~C.,  {Somerville} R.~S.,  {Primack} J.~R.,
   {Dekel} A.,  {Rodr{\'\i}guez-Puebla} A.,  {Guo} Y.,  {Barro} G.,  {Kocevski}
  D.~D.,  {van der Wel} A.,  {Woo} J.,  {Bell} E.~F.,  {Fang} J.~J.,    et al.
  2020, \apj, 897, 102

\bibitem[\protect\citeauthoryear{{Cheung}, {Faber}, {Koo}, {Dutton}, {Simard},
  {McGrath}, {Huang}, {Bell}, {Dekel}, {Fang}, {Salim}, {Barro}, {Bundy} \& et
  al.}{{Cheung} et~al.}{2012}]{Cheung2012}
{Cheung} E.,  {Faber} S.~M.,  {Koo} D.~C.,  {Dutton} A.~A.,  {Simard} L.,
  {McGrath} E.~J.,  {Huang} J.~S.,  {Bell} E.~F.,  {Dekel} A.,  {Fang} J.~J.,
  {Salim} S.,  {Barro} G.,  {Bundy} K.,    et al. 2012, \apj, 760, 131

\bibitem[\protect\citeauthoryear{{Christensen}, {Dav{\'e}}, {Governato},
  {Pontzen}, {Brooks}, {Munshi}, {Quinn} \& {Wadsley}}{{Christensen}
  et~al.}{2016}]{Christensen2016}
{Christensen} C.~R.,  {Dav{\'e}} R.,  {Governato} F.,  {Pontzen} A.,  {Brooks}
  A.,  {Munshi} F.,  {Quinn} T.,    {Wadsley} J.,  2016, \apj, 824, 57

\bibitem[\protect\citeauthoryear{{Combes}, {Garc{\'\i}a-Burillo}, {Braine},
  {Schinnerer}, {Walter} \& {Colina}}{{Combes} et~al.}{2013}]{Combes2013}
{Combes} F.,  {Garc{\'\i}a-Burillo} S.,  {Braine} J.,  {Schinnerer} E.,
  {Walter} F.,    {Colina} L.,  2013, \aap, 550, A41

\bibitem[\protect\citeauthoryear{{Cora}, {Vega-Mart{\'\i}nez}, {Hough}, {Ruiz},
  {Orsi}, {Mu{\~n}oz Arancibia}, {Gargiulo}, {Collacchioni}, {Padilla},
  {Gottl{\"o}ber} \& {Yepes}}{{Cora} et~al.}{2018}]{Cora2018}
{Cora} S.~A.,  {Vega-Mart{\'\i}nez} C.~A.,  {Hough} T.,  {Ruiz} A.~N.,  {Orsi}
  {\'A}.~A.,  {Mu{\~n}oz Arancibia} A.~M.,  {Gargiulo} I.~D.,  {Collacchioni}
  F.,  {Padilla} N.~D.,  {Gottl{\"o}ber} S.,    {Yepes} G.,  2018, \mnras, 479,
  2

\bibitem[\protect\citeauthoryear{{Croton}, {Springel}, {White}, {De Lucia},
  {Frenk}, {Gao}, {Jenkins}, {Kauffmann}, {Navarro} \& {Yoshida}}{{Croton}
  et~al.}{2006}]{Croton2006}
{Croton} D.~J.,  {Springel} V.,  {White} S. D.~M.,  {De Lucia} G.,  {Frenk}
  C.~S.,  {Gao} L.,  {Jenkins} A.,  {Kauffmann} G.,  {Navarro} J.~F.,
  {Yoshida} N.,  2006, \mnras, 365, 11

\bibitem[\protect\citeauthoryear{{Croton}, {Stevens}, {Tonini}, {Garel},
  {Bernyk}, {Bibiano}, {Hodkinson}, {Mutch}, {Poole} \& {Shattow}}{{Croton}
  et~al.}{2016}]{Croton2016}
{Croton} D.~J.,  {Stevens} A. R.~H.,  {Tonini} C.,  {Garel} T.,  {Bernyk} M.,
  {Bibiano} A.,  {Hodkinson} L.,  {Mutch} S.~J.,  {Poole} G.~B.,    {Shattow}
  G.~M.,  2016, \apjs, 222, 22

\bibitem[\protect\citeauthoryear{{Davies}, {Crain}, {McCarthy}, {Oppenheimer},
  {Schaye}, {Schaller} \& {McAlpine}}{{Davies} et~al.}{2019}]{Davies2019b}
{Davies} J.~J.,  {Crain} R.~A.,  {McCarthy} I.~G.,  {Oppenheimer} B.~D.,
  {Schaye} J.,  {Schaller} M.,    {McAlpine} S.,  2019, \mnras, 485, 3783

\bibitem[\protect\citeauthoryear{{Davies}, {Crain}, {Oppenheimer} \&
  {Schaye}}{{Davies} et~al.}{2020}]{Davies2020}
{Davies} J.~J.,  {Crain} R.~A.,  {Oppenheimer} B.~D.,    {Schaye} J.,  2020,
  \mnras, 491, 4462

\bibitem[\protect\citeauthoryear{{Davies}, {Robotham}, {Lagos}, {Driver},
  {Stevens}, {Bah{\'e}}, {Alpaslan}, {Bremer}, {Brown}, {Brough},
  {Bland-Hawthorn} \& et al.}{{Davies} et~al.}{2019}]{Davies2019}
{Davies} L.~J.~M.,  {Robotham} A.~S.~G.,  {Lagos} C. d.~P.,  {Driver} S.~P.,
  {Stevens} A.~R.~H.,  {Bah{\'e}} Y.~M.,  {Alpaslan} M.,  {Bremer} M.~N.,
  {Brown} M.~J.~I.,  {Brough} S.,  {Bland-Hawthorn} J.,    et al. 2019, \mnras,
  483, 5444

\bibitem[\protect\citeauthoryear{{Davis}, {Graham} \& {Cameron}}{{Davis}
  et~al.}{2018}]{Davis2018}
{Davis} B.~L.,  {Graham} A.~W.,    {Cameron} E.,  2018, \apj, 869, 113

\bibitem[\protect\citeauthoryear{{de Vaucouleurs}}{{de
  Vaucouleurs}}{1948}]{deVaucouleurs1948}
{de Vaucouleurs} G.,  1948, Annales d'Astrophysique, 11, 247

\bibitem[\protect\citeauthoryear{{Dekel} \& {Birnboim}}{{Dekel} \&
  {Birnboim}}{2006}]{Dekel2006}
{Dekel} A.,  {Birnboim} Y.,  2006, \mnras, 368, 2

\bibitem[\protect\citeauthoryear{{Dekel}, {Lapiner} \& {Dubois}}{{Dekel}
  et~al.}{2019}]{Dekel2019}
{Dekel} A.,  {Lapiner} S.,    {Dubois} Y.,  2019, arXiv e-prints, p.
  arXiv:1904.08431

\bibitem[\protect\citeauthoryear{{Delvecchio}, {Lutz}, {Berta}, {Rosario},
  {Zamorani}, {Pozzi}, {Gruppioni}, {Vignali}, {Brusa}, {Cimatti}, {Clements}
  \& et al.}{{Delvecchio} et~al.}{2015}]{Delvecchio2015}
{Delvecchio} I.,  {Lutz} D.,  {Berta} S.,  {Rosario} D.~J.,  {Zamorani} G.,
  {Pozzi} F.,  {Gruppioni} C.,  {Vignali} C.,  {Brusa} M.,  {Cimatti} A.,
  {Clements} D.~L.,    et al. 2015, \mnras, 449, 373

\bibitem[\protect\citeauthoryear{{Devergne}, {Cattaneo}, {Bournaud},
  {Koutsouridou}, {Winter}, {Dimauro}, {Mamon}, {Vacher} \& {Varin}}{{Devergne}
  et~al.}{2020}]{Devergne2020}
{Devergne} T.,  {Cattaneo} A.,  {Bournaud} F.,  {Koutsouridou} I.,  {Winter}
  A.,  {Dimauro} P.,  {Mamon} G.~A.,  {Vacher} W.,    {Varin} M.,  2020, \aap,
  644, A56

\bibitem[\protect\citeauthoryear{{Dimauro}, {Huertas-Company}, {Daddi},
  {P{\'e}rez-Gonz{\'a}lez}, {Bernardi}, {Barro}, {Buitrago}, {Caro},
  {Cattaneo}, {Dominguez-S{\'a}nchez}, {Faber}, {H{\"a}u{\ss}ler} \& et
  al.}{{Dimauro} et~al.}{2018}]{Dimauro2018}
{Dimauro} P.,  {Huertas-Company} M.,  {Daddi} E.,  {P{\'e}rez-Gonz{\'a}lez}
  P.~G.,  {Bernardi} M.,  {Barro} G.,  {Buitrago} F.,  {Caro} F.,  {Cattaneo}
  A.,  {Dominguez-S{\'a}nchez} H.,  {Faber} S. r.~M.,  {H{\"a}u{\ss}ler} B.,
  et al. 2018, \mnras, 478, 5410

\bibitem[\protect\citeauthoryear{{Dubois}, {Volonteri}, {Silk}, {Devriendt},
  {Slyz} \& {Teyssier}}{{Dubois} et~al.}{2015}]{Dubois2015}
{Dubois} Y.,  {Volonteri} M.,  {Silk} J.,  {Devriendt} J.,  {Slyz} A.,
  {Teyssier} R.,  2015, \mnras, 452, 1502

\bibitem[\protect\citeauthoryear{{Duc} \& {Bournaud}}{{Duc} \&
  {Bournaud}}{2008}]{Duc2008}
{Duc} P.-A.,  {Bournaud} F.,  2008, \apj, 673, 787

\bibitem[\protect\citeauthoryear{{Ebeling}, {Stephenson} \& {Edge}}{{Ebeling}
  et~al.}{2014}]{Ebeling2014}
{Ebeling} H.,  {Stephenson} L.~N.,    {Edge} A.~C.,  2014, \apjl, 781, L40

\bibitem[\protect\citeauthoryear{{Edge} \& {Stewart}}{{Edge} \&
  {Stewart}}{1991}]{edge_stewart91}
{Edge} A.~C.,  {Stewart} G.~C.,  1991, \mnras, 252, 414

\bibitem[\protect\citeauthoryear{{Efstathiou}}{{Efstathiou}}{1992}]{Efstathiou1992}
{Efstathiou} G.,  1992, \mnras, 256, 43P

\bibitem[\protect\citeauthoryear{{Efstathiou}, {Lake} \&
  {Negroponte}}{{Efstathiou} et~al.}{1982}]{Efstathiou1982}
{Efstathiou} G.,  {Lake} G.,    {Negroponte} J.,  1982, \mnras, 199, 1069

\bibitem[\protect\citeauthoryear{{Evrard} \& {Henry}}{{Evrard} \&
  {Henry}}{1991}]{evrard_henry91}
{Evrard} A.~E.,  {Henry} J.~P.,  1991, \apj, 383, 95

\bibitem[\protect\citeauthoryear{{Fabian}}{{Fabian}}{1999}]{Fabian1999}
{Fabian} A.~C.,  1999, \mnras, 308, L39

\bibitem[\protect\citeauthoryear{{Faltenbacher}, {Hoffman}, {Gottl{\"o}ber} \&
  {Yepes}}{{Faltenbacher} et~al.}{2007}]{Faltenbacher2007}
{Faltenbacher} A.,  {Hoffman} Y.,  {Gottl{\"o}ber} S.,    {Yepes} G.,  2007,
  \mnras, 376, 1327

\bibitem[\protect\citeauthoryear{{Fan}}{{Fan}}{2006}]{Fan2006}
{Fan} X.,  2006, \nar, 50, 665

\bibitem[\protect\citeauthoryear{{Fanaroff} \& {Riley}}{{Fanaroff} \&
  {Riley}}{1974}]{fanaroff_riley74}
{Fanaroff} B.~L.,  {Riley} J.~M.,  1974, \mnras, 167, 31P

\bibitem[\protect\citeauthoryear{{Fang}, {Faber}, {Koo} \& {Dekel}}{{Fang}
  et~al.}{2013}]{Fang2013}
{Fang} J.~J.,  {Faber} S.~M.,  {Koo} D.~C.,    {Dekel} A.,  2013, \apj, 776, 63

\bibitem[\protect\citeauthoryear{{Farouki} \& {Shapiro}}{{Farouki} \&
  {Shapiro}}{1981}]{Farouki1981}
{Farouki} R.,  {Shapiro} S.~L.,  1981, \apj, 243, 32

\bibitem[\protect\citeauthoryear{{Ferrarese} \& {Merritt}}{{Ferrarese} \&
  {Merritt}}{2000}]{Ferrarese2000}
{Ferrarese} L.,  {Merritt} D.,  2000, \apjl, 539, L9

\bibitem[\protect\citeauthoryear{{Font}, {Bower}, {McCarthy}, {Benson},
  {Frenk}, {Helly}, {Lacey}, {Baugh} \& {Cole}}{{Font} et~al.}{2008}]{Font2008}
{Font} A.~S.,  {Bower} R.~G.,  {McCarthy} I.~G.,  {Benson} A.~J.,  {Frenk}
  C.~S.,  {Helly} J.~C.,  {Lacey} C.~G.,  {Baugh} C.~M.,    {Cole} S.,  2008,
  \mnras, 389, 1619

\bibitem[\protect\citeauthoryear{{Gargiulo}, {Cora}, {Padilla}, {Mu{\~n}oz
  Arancibia}, {Ruiz}, {Orsi}, {Tecce}, {Weidner} \& {Bruzual}}{{Gargiulo}
  et~al.}{2015}]{Gargiulo2015}
{Gargiulo} I.~D.,  {Cora} S.~A.,  {Padilla} N.~D.,  {Mu{\~n}oz Arancibia}
  A.~M.,  {Ruiz} A.~N.,  {Orsi} A.~A.,  {Tecce} T.~E.,  {Weidner} C.,
  {Bruzual} G.,  2015, \mnras, 446, 3820

\bibitem[\protect\citeauthoryear{{Gebhardt}, {Bender}, {Bower}, {Dressler},
  {Faber}, {Filippenko}, {Green}, {Grillmair}, {Ho}, {Kormendy}, {Lauer},
  {Magorrian}, {Pinkney}, {Richstone} \& {Tremaine}}{{Gebhardt}
  et~al.}{2000}]{Gebhardt2000}
{Gebhardt} K.,  {Bender} R.,  {Bower} G.,  {Dressler} A.,  {Faber} S.~M.,
  {Filippenko} A.~V.,  {Green} R.,  {Grillmair} C.,  {Ho} L.~C.,  {Kormendy}
  J.,  {Lauer} T.~R.,  {Magorrian} J.,  {Pinkney} J.,  {Richstone} D.,
  {Tremaine} S.,  2000, \apjl, 539, L13

\bibitem[\protect\citeauthoryear{{Gnedin}}{{Gnedin}}{2000}]{Gnedin2000}
{Gnedin} N.~Y.,  2000, \apj, 542, 535

\bibitem[\protect\citeauthoryear{{Gonzalez-Perez}, {Lacey}, {Baugh}, {Lagos},
  {Helly}, {Campbell} \& {Mitchell}}{{Gonzalez-Perez}
  et~al.}{2014}]{GonzalezPerez2014}
{Gonzalez-Perez} V.,  {Lacey} C.~G.,  {Baugh} C.~M.,  {Lagos} C.~D.~P.,
  {Helly} J.,  {Campbell} D.~J.~R.,    {Mitchell} P.~D.,  2014, \mnras, 439,
  264

\bibitem[\protect\citeauthoryear{{Gunn} \& {Gott} III}{{Gunn} \&
  {Gott}}{1972}]{Gunn1972}
{Gunn} J.~E.,  {Gott} III J.~R.,  1972, \apj, 176, 1

\bibitem[\protect\citeauthoryear{{Guo}, {Cole}, {Eke}, {Frenk} \&
  {Helly}}{{Guo} et~al.}{2013}]{Guo2013}
{Guo} Q.,  {Cole} S.,  {Eke} V.,  {Frenk} C.,    {Helly} J.,  2013, \mnras,
  434, 1838

\bibitem[\protect\citeauthoryear{{Guo}, {White}, {Boylan-Kolchin}, {De Lucia},
  {Kauffmann}, {Lemson}, {Li}, {Springel} \& {Weinmann}}{{Guo}
  et~al.}{2011}]{Guo2011}
{Guo} Q.,  {White} S.,  {Boylan-Kolchin} M.,  {De Lucia} G.,  {Kauffmann} G.,
  {Lemson} G.,  {Li} C.,  {Springel} V.,    {Weinmann} S.,  2011, \mnras, 413,
  101

\bibitem[\protect\citeauthoryear{{Habouzit}, {Volonteri} \&
  {Dubois}}{{Habouzit} et~al.}{2017}]{Habouzit2017}
{Habouzit} M.,  {Volonteri} M.,    {Dubois} Y.,  2017, \mnras, 468, 3935

\bibitem[\protect\citeauthoryear{{Haemmerl{\'e}}, {Mayer}, {Klessen},
  {Hosokawa}, {Madau} \& {Bromm}}{{Haemmerl{\'e}} et~al.}{2020}]{Haemmerle2020}
{Haemmerl{\'e}} L.,  {Mayer} L.,  {Klessen} R.~S.,  {Hosokawa} T.,  {Madau} P.,
     {Bromm} V.,  2020, \ssr, 216, 48

\bibitem[\protect\citeauthoryear{{Harris}, {Farrah}, {Schulz},
  {Hatziminaoglou}, {Viero}, {Anderson}, {B{\'e}thermin}, {Chapman},
  {Clements}, {Cooray} \& et al.}{{Harris} et~al.}{2016}]{Harris2016}
{Harris} K.,  {Farrah} D.,  {Schulz} B.,  {Hatziminaoglou} E.,  {Viero} M.,
  {Anderson} N.,  {B{\'e}thermin} M.,  {Chapman} S.,  {Clements} D.~L.,
  {Cooray} A.,    et al. 2016, \mnras, 457, 4179

\bibitem[\protect\citeauthoryear{{Hatton}, {Devriendt}, {Ninin}, {Bouchet},
  {Guiderdoni} \& {Vibert}}{{Hatton} et~al.}{2003}]{Hatton2003}
{Hatton} S.,  {Devriendt} J. E.~G.,  {Ninin} S.,  {Bouchet} F.~R.,
  {Guiderdoni} B.,    {Vibert} D.,  2003, \mnras, 343, 75

\bibitem[\protect\citeauthoryear{{Henriques}, {White}, {Thomas}, {Angulo},
  {Guo}, {Lemson}, {Springel} \& {Overzier}}{{Henriques}
  et~al.}{2015}]{Henriques2015}
{Henriques} B. M.~B.,  {White} S. D.~M.,  {Thomas} P.~A.,  {Angulo} R.,  {Guo}
  Q.,  {Lemson} G.,  {Springel} V.,    {Overzier} R.,  2015, \mnras, 451, 2663

\bibitem[\protect\citeauthoryear{{Hlavacek-Larrondo}, {Li} \&
  {Churazov}}{{Hlavacek-Larrondo} et~al.}{2022}]{HlavecekLarrondo2022}
{Hlavacek-Larrondo} J.,  {Li} Y.,    {Churazov} E.,  2022, arXiv e-prints, p.
  arXiv:2206.00098

\bibitem[\protect\citeauthoryear{{Hopkins}, {Cox}, {Younger} \&
  {Hernquist}}{{Hopkins} et~al.}{2009}]{Hopkins2009}
{Hopkins} P.~F.,  {Cox} T.~J.,  {Younger} J.~D.,    {Hernquist} L.,  2009,
  \apj, 691, 1168

\bibitem[\protect\citeauthoryear{{Hopkins}, {Hernquist}, {Cox}, {Di Matteo},
  {Robertson} \& {Springel}}{{Hopkins} et~al.}{2006}]{Hopkins2006}
{Hopkins} P.~F.,  {Hernquist} L.,  {Cox} T.~J.,  {Di Matteo} T.,  {Robertson}
  B.,    {Springel} V.,  2006, \apjs, 163, 1

\bibitem[\protect\citeauthoryear{{Hopkins}, {Hernquist}, {Cox} \&
  {Kere{\v{s}}}}{{Hopkins} et~al.}{2008}]{Hopkins2008a}
{Hopkins} P.~F.,  {Hernquist} L.,  {Cox} T.~J.,    {Kere{\v{s}}} D.,  2008,
  \apjs, 175, 356

\bibitem[\protect\citeauthoryear{{Hopkins}, {Hernquist}, {Cox}, {Younger} \&
  {Besla}}{{Hopkins} et~al.}{2008}]{Hopkins2008b}
{Hopkins} P.~F.,  {Hernquist} L.,  {Cox} T.~J.,  {Younger} J.~D.,    {Besla}
  G.,  2008, \apj, 688, 757

\bibitem[\protect\citeauthoryear{{Hou}, {Li}, {Jones}, {Forman} \& {Su}}{{Hou}
  et~al.}{2021}]{hou_etal21}
{Hou} M.,  {Li} Z.,  {Jones} C.,  {Forman} W.,    {Su} Y.,  2021, \apj, 919,
  141

\bibitem[\protect\citeauthoryear{{Ilbert}, {McCracken}, {Le F{\`e}vre},
  {Capak}, {Dunlop}, {Karim}, {Renzini}, {Caputi}, {Boissier}, {Arnouts},
  {Aussel}, {Comparat}, {Guo}, {Hudelot}, {Kartaltepe} \& et al.}{{Ilbert}
  et~al.}{2013}]{Ilbert2013}
{Ilbert} O.,  {McCracken} H.~J.,  {Le F{\`e}vre} O.,  {Capak} P.,  {Dunlop} J.,
   {Karim} A.,  {Renzini} M.~A.,  {Caputi} K.,  {Boissier} S.,  {Arnouts} S.,
  {Aussel} H.,  {Comparat} J.,  {Guo} Q.,  {Hudelot} P.,  {Kartaltepe} J.,
  et al. 2013, \aap, 556, A55

\bibitem[\protect\citeauthoryear{{Inayoshi}, {Visbal} \& {Haiman}}{{Inayoshi}
  et~al.}{2019}]{Inayoshi2019}
{Inayoshi} K.,  {Visbal} E.,    {Haiman} Z.,  2019, arXiv e-prints, p.
  arXiv:1911.05791

\bibitem[\protect\citeauthoryear{{Jiang}, {Dekel}, {Kneller}, {Lapiner},
  {Ceverino}, {Primack}, {Faber}, {Macci{\`o}}, {Dutton}, {Genel} \&
  {Somerville}}{{Jiang} et~al.}{2019}]{Jiang2019}
{Jiang} F.,  {Dekel} A.,  {Kneller} O.,  {Lapiner} S.,  {Ceverino} D.,
  {Primack} J.~R.,  {Faber} S.~M.,  {Macci{\`o}} A.~V.,  {Dutton} A.~A.,
  {Genel} S.,    {Somerville} R.~S.,  2019, \mnras, 488, 4801

\bibitem[\protect\citeauthoryear{{Kaiser}}{{Kaiser}}{1991}]{kaiser91}
{Kaiser} N.,  1991, \apj, 383, 104

\bibitem[\protect\citeauthoryear{{Kang} \& {van den Bosch}}{{Kang} \& {van den
  Bosch}}{2008}]{Kang2008}
{Kang} X.,  {van den Bosch} F.~C.,  2008, \apjl, 676, L101

\bibitem[\protect\citeauthoryear{{Kannan}, {Macci{\`o}}, {Fontanot}, {Moster},
  {Karman} \& {Somerville}}{{Kannan} et~al.}{2015}]{Kannan2015}
{Kannan} R.,  {Macci{\`o}} A.~V.,  {Fontanot} F.,  {Moster} B.~P.,  {Karman}
  W.,    {Somerville} R.~S.,  2015, \mnras, 452, 4347

\bibitem[\protect\citeauthoryear{{Kauffmann} \& {Haehnelt}}{{Kauffmann} \&
  {Haehnelt}}{2000}]{Kauffmann2000}
{Kauffmann} G.,  {Haehnelt} M.,  2000, \mnras, 311, 576

\bibitem[\protect\citeauthoryear{{Kauffmann}, {Heckman}, {Tremonti},
  {Brinchmann}, {Charlot}, {White}, {Ridgway}, {Brinkmann}, {Fukugita}, {Hall}
  \& et al.}{{Kauffmann} et~al.}{2003}]{Kauffmann2003}
{Kauffmann} G.,  {Heckman} T.~M.,  {Tremonti} C.,  {Brinchmann} J.,  {Charlot}
  S.,  {White} S. D.~M.,  {Ridgway} S.~E.,  {Brinkmann} J.,  {Fukugita} M.,
  {Hall} P.~B.,    et al. 2003, \mnras, 346, 1055

\bibitem[\protect\citeauthoryear{{Kennicutt}
  Robert~C.}{{Kennicutt}}{1998}]{Kennicutt1998}
{Kennicutt} Robert~C. J.,  1998, \apj, 498, 541

\bibitem[\protect\citeauthoryear{{Kennicutt} Robert~C. \& {De Los
  Reyes}}{{Kennicutt} \& {De Los Reyes}}{2021}]{Kennicutt2021}
{Kennicutt} Robert~C. J.,  {De Los Reyes} M. A.~C.,  2021, \apj, 908, 61

\bibitem[\protect\citeauthoryear{{Kennicutt} \& {Evans}}{{Kennicutt} \&
  {Evans}}{2012}]{Kennicutt2012}
{Kennicutt} R.~C.,  {Evans} N.~J.,  2012, \araa, 50, 531

\bibitem[\protect\citeauthoryear{{Kimm}, {Somerville}, {Yi}, {van den Bosch},
  {Salim}, {Fontanot}, {Monaco}, {Mo}, {Pasquali}, {Rich} \& {Yang}}{{Kimm}
  et~al.}{2009}]{Kimm2009}
{Kimm} T.,  {Somerville} R.~S.,  {Yi} S.~K.,  {van den Bosch} F.~C.,  {Salim}
  S.,  {Fontanot} F.,  {Monaco} P.,  {Mo} H.,  {Pasquali} A.,  {Rich} R.~M.,
  {Yang} X.,  2009, \mnras, 394, 1131

\bibitem[\protect\citeauthoryear{{King}}{{King}}{2003}]{King2003}
{King} A.,  2003, \apjl, 596, L27

\bibitem[\protect\citeauthoryear{{Knebe}, {Pearce}, {Thomas}, {Benson},
  {Blaizot}, {Bower}, {Carretero}, {Castander}, {Cattaneo}, {Cora}, {Croton},
  {Cui}, {Cunnama}, {De Lucia}, {Devriendt} \& et al.}{{Knebe}
  et~al.}{2015}]{Knebe2015}
{Knebe} A.,  {Pearce} F.~R.,  {Thomas} P.~A.,  {Benson} A.,  {Blaizot} J.,
  {Bower} R.,  {Carretero} J.,  {Castander} F.~J.,  {Cattaneo} A.,  {Cora}
  S.~A.,  {Croton} D.~J.,  {Cui} W.,  {Cunnama} D.,  {De Lucia} G.,
  {Devriendt} J.~E.,    et al. 2015, \mnras, 451, 4029

\bibitem[\protect\citeauthoryear{{Kormendy}, {Fisher}, {Cornell} \&
  {Bender}}{{Kormendy} et~al.}{2009}]{Kormendy2009}
{Kormendy} J.,  {Fisher} D.~B.,  {Cornell} M.~E.,    {Bender} R.,  2009, \apjs,
  182, 216

\bibitem[\protect\citeauthoryear{{Koutsouridou} \& {Cattaneo}}{{Koutsouridou}
  \& {Cattaneo}}{2019}]{Koutsouridou2019}
{Koutsouridou} I.,  {Cattaneo} A.,  2019, \mnras, 490, 5375

\bibitem[\protect\citeauthoryear{{Krumholz}, {Dekel} \& {McKee}}{{Krumholz}
  et~al.}{2012}]{Krumholz2012}
{Krumholz} M.~R.,  {Dekel} A.,    {McKee} C.~F.,  2012, \apj, 745, 69

\bibitem[\protect\citeauthoryear{{Lagos}, {Tobar}, {Robotham}, {Obreschkow},
  {Mitchell}, {Power} \& {Elahi}}{{Lagos} et~al.}{2018}]{Lagos2018}
{Lagos} C. d.~P.,  {Tobar} R.~J.,  {Robotham} A. S.~G.,  {Obreschkow} D.,
  {Mitchell} P.~D.,  {Power} C.,    {Elahi} P.~J.,  2018, \mnras, 481, 3573

\bibitem[\protect\citeauthoryear{{Lamastra}, {Menci}, {Fiore} \&
  {Santini}}{{Lamastra} et~al.}{2013}]{Lamastra2013}
{Lamastra} A.,  {Menci} N.,  {Fiore} F.,    {Santini} P.,  2013, \aap, 552, A44

\bibitem[\protect\citeauthoryear{{Larson}, {Tinsley} \& {Caldwell}}{{Larson}
  et~al.}{1980}]{Larson1980}
{Larson} R.~B.,  {Tinsley} B.~M.,    {Caldwell} C.~N.,  1980, \apj, 237, 692

\bibitem[\protect\citeauthoryear{{Latif} \& {Ferrara}}{{Latif} \&
  {Ferrara}}{2016}]{Latif2016}
{Latif} M.~A.,  {Ferrara} A.,  2016, \pasa, 33, e051

\bibitem[\protect\citeauthoryear{{Lee} \& {Yi}}{{Lee} \& {Yi}}{2013}]{Lee2013}
{Lee} J.,  {Yi} S.~K.,  2013, \apj, 766, 38

\bibitem[\protect\citeauthoryear{{Lloyd-Davies}, {Ponman} \&
  {Cannon}}{{Lloyd-Davies} et~al.}{2000}]{lloyd_etal00}
{Lloyd-Davies} E.~J.,  {Ponman} T.~J.,    {Cannon} D.~B.,  2000, \mnras, 315,
  689

\bibitem[\protect\citeauthoryear{{Lo Faro}, {Monaco}, {Vanzella}, {Fontanot},
  {Silva} \& {Cristiani}}{{Lo Faro} et~al.}{2009}]{LoFaro2009}
{Lo Faro} B.,  {Monaco} P.,  {Vanzella} E.,  {Fontanot} F.,  {Silva} L.,
  {Cristiani} S.,  2009, \mnras, 399, 827

\bibitem[\protect\citeauthoryear{{Mamon}, {Trevisan}, {Thuan}, {Gallazzi} \&
  {Dav{\'e}}}{{Mamon} et~al.}{2020}]{Mamon2020}
{Mamon} G.~A.,  {Trevisan} M.,  {Thuan} T.~X.,  {Gallazzi} A.,    {Dav{\'e}}
  R.,  2020, \mnras, 492, 1791

\bibitem[\protect\citeauthoryear{{Martin}, {Kaviraj}, {Devriendt}, {Dubois} \&
  {Pichon}}{{Martin} et~al.}{2018}]{Martin2018}
{Martin} G.,  {Kaviraj} S.,  {Devriendt} J.~E.~G.,  {Dubois} Y.,    {Pichon}
  C.,  2018, \mnras, 480, 2266

\bibitem[\protect\citeauthoryear{{Mayer}, {Kazantzidis}, {Escala} \&
  {Callegari}}{{Mayer} et~al.}{2010}]{Mayer2010}
{Mayer} L.,  {Kazantzidis} S.,  {Escala} A.,    {Callegari} S.,  2010, \nat,
  466, 1082

\bibitem[\protect\citeauthoryear{{Mayer}, {Mastropietro}, {Wadsley}, {Stadel}
  \& {Moore}}{{Mayer} et~al.}{2006}]{Mayer2006}
{Mayer} L.,  {Mastropietro} C.,  {Wadsley} J.,  {Stadel} J.,    {Moore} B.,
  2006, \mnras, 369, 1021

\bibitem[\protect\citeauthoryear{{McCarthy}, {Frenk}, {Font}, {Lacey}, {Bower},
  {Mitchell}, {Balogh} \& {Theuns}}{{McCarthy} et~al.}{2008}]{McCarthy2008}
{McCarthy} I.~G.,  {Frenk} C.~S.,  {Font} A.~S.,  {Lacey} C.~G.,  {Bower}
  R.~G.,  {Mitchell} N.~L.,  {Balogh} M.~L.,    {Theuns} T.,  2008, \mnras,
  383, 593

\bibitem[\protect\citeauthoryear{{McNamara} \& {Nulsen}}{{McNamara} \&
  {Nulsen}}{2007}]{Mcnamara2007}
{McNamara} B.~R.,  {Nulsen} P.~E.~J.,  2007, \araa, 45, 117

\bibitem[\protect\citeauthoryear{{McNamara} \& {Nulsen}}{{McNamara} \&
  {Nulsen}}{2012}]{Mcnamara2012}
{McNamara} B.~R.,  {Nulsen} P.~E.~J.,  2012, New Journal of Physics, 14, 055023

\bibitem[\protect\citeauthoryear{{Mendel}, {Simard}, {Ellison} \&
  {Patton}}{{Mendel} et~al.}{2013}]{Mendel2013}
{Mendel} J.~T.,  {Simard} L.,  {Ellison} S.~L.,    {Patton} D.~R.,  2013,
  \mnras, 429, 2212

\bibitem[\protect\citeauthoryear{{Mendel}, {Simard}, {Palmer}, {Ellison} \&
  {Patton}}{{Mendel} et~al.}{2014}]{Mendel2014}
{Mendel} J.~T.,  {Simard} L.,  {Palmer} M.,  {Ellison} S.~L.,    {Patton}
  D.~R.,  2014, \apjs, 210, 3

\bibitem[\protect\citeauthoryear{{Merritt}}{{Merritt}}{1984}]{Merritt1984}
{Merritt} D.,  1984, \apj, 276, 26

\bibitem[\protect\citeauthoryear{{Mo}, {Mao} \& {White}}{{Mo}
  et~al.}{1998}]{Mo1998}
{Mo} H.~J.,  {Mao} S.,    {White} S. D.~M.,  1998, \mnras, 295, 319

\bibitem[\protect\citeauthoryear{{Moore}, {Katz}, {Lake}, {Dressler} \&
  {Oemler}}{{Moore} et~al.}{1996}]{Moore1996}
{Moore} B.,  {Katz} N.,  {Lake} G.,  {Dressler} A.,    {Oemler} A.,  1996,
  \nat, 379, 613

\bibitem[\protect\citeauthoryear{{Morselli}, {Popesso}, {Erfanianfar} \&
  {Concas}}{{Morselli} et~al.}{2017}]{Morselli2017}
{Morselli} L.,  {Popesso} P.,  {Erfanianfar} G.,    {Concas} A.,  2017, \aap,
  597, A97

\bibitem[\protect\citeauthoryear{{Moustakas}, {Coil}, {Aird}, {Blanton},
  {Cool}, {Eisenstein}, {Mendez}, {Wong}, {Zhu} \& {Arnouts}}{{Moustakas}
  et~al.}{2013}]{Moustakas2013}
{Moustakas} J.,  {Coil} A.~L.,  {Aird} J.,  {Blanton} M.~R.,  {Cool} R.~J.,
  {Eisenstein} D.~J.,  {Mendez} A.~J.,  {Wong} K.~C.,  {Zhu} G.,    {Arnouts}
  S.,  2013, \apj, 767, 50

\bibitem[\protect\citeauthoryear{{Muzzin}, {Marchesini}, {Stefanon}, {Franx},
  {McCracken}, {Milvang-Jensen}, {Dunlop}, {Fynbo}, {Brammer}, {Labb{\'e}} \&
  {van Dokkum}}{{Muzzin} et~al.}{2013}]{Muzzin2013}
{Muzzin} A.,  {Marchesini} D.,  {Stefanon} M.,  {Franx} M.,  {McCracken} H.~J.,
   {Milvang-Jensen} B.,  {Dunlop} J.~S.,  {Fynbo} J.~P.~U.,  {Brammer} G.,
  {Labb{\'e}} I.,    {van Dokkum} P.~G.,  2013, \apj, 777, 18

\bibitem[\protect\citeauthoryear{{Navarro}, {Frenk} \& {White}}{{Navarro}
  et~al.}{1997}]{Navarro1997}
{Navarro} J.~F.,  {Frenk} C.~S.,    {White} S. D.~M.,  1997, \apj, 490, 493

\bibitem[\protect\citeauthoryear{{Negroponte} \& {White}}{{Negroponte} \&
  {White}}{1983}]{Negroponte1983}
{Negroponte} J.,  {White} S.~D.~M.,  1983, \mnras, 205, 1009

\bibitem[\protect\citeauthoryear{{Oppenheimer}}{{Oppenheimer}}{2018}]{Oppenheimer2018}
{Oppenheimer} B.~D.,  2018, \mnras, 480, 2963

\bibitem[\protect\citeauthoryear{{Oppenheimer}, {Davies}, {Crain}, {Wijers},
  {Schaye}, {Werk}, {Burchett}, {Trayford} \& {Horton}}{{Oppenheimer}
  et~al.}{2020}]{Oppenheimer2020}
{Oppenheimer} B.~D.,  {Davies} J.~J.,  {Crain} R.~A.,  {Wijers} N.~A.,
  {Schaye} J.,  {Werk} J.~K.,  {Burchett} J.~N.,  {Trayford} J.~W.,    {Horton}
  R.,  2020, \mnras, 491, 2939

\bibitem[\protect\citeauthoryear{{Ostriker}, {Bode} \& {Babul}}{{Ostriker}
  et~al.}{2005}]{Ostriker2005}
{Ostriker} J.~P.,  {Bode} P.,    {Babul} A.,  2005, \apj, 634, 964

\bibitem[\protect\citeauthoryear{{Pan}, {Lin}, {Hsieh}, {Barrera-Ballesteros},
  {S{\'a}nchez}, {Hsu}, {Keenan}, {Tissera}, {Boquien}, {Dai}, {Knapen},
  {Riffel}, {Argudo-Fern{\'a}ndez}, {Xiao} \& {Yuan}}{{Pan}
  et~al.}{2019}]{Pan2019}
{Pan} H.-A.,  {Lin} L.,  {Hsieh} B.-C.,  {Barrera-Ballesteros} J.~K.,
  {S{\'a}nchez} S.~F.,  {Hsu} C.-H.,  {Keenan} R.,  {Tissera} P.~B.,  {Boquien}
  M.,  {Dai} Y.~S.,  {Knapen} J.~H.,  {Riffel} R.,  {Argudo-Fern{\'a}ndez} M.,
  {Xiao} T.,    {Yuan} F.-T.,  2019, \apj, 881, 119

\bibitem[\protect\citeauthoryear{{Peterson} \& {Fabian}}{{Peterson} \&
  {Fabian}}{2006}]{Peterson2006}
{Peterson} J.~R.,  {Fabian} A.~C.,  2006, \physrep, 427, 1

\bibitem[\protect\citeauthoryear{{Planck Collaboration}, {Ade}, {Aghanim},
  {Armitage-Caplan}, {Arnaud}, {Ashdown}, {Atrio-Barandela}, {Aumont},
  {Baccigalupi}, {Banday} \& et al.}{{Planck Collaboration}
  et~al.}{2014}]{planck_etal14}
{Planck Collaboration} {Ade} P.~A.~R.,  {Aghanim} N.,  {Armitage-Caplan} C.,
  {Arnaud} M.,  {Ashdown} M.,  {Atrio-Barandela} F.,  {Aumont} J.,
  {Baccigalupi} C.,  {Banday} A.~J.,    et al. 2014, \aap, 571, A20

\bibitem[\protect\citeauthoryear{{Ponman}, {Bourner}, {Ebeling} \&
  {B{\"o}hringer}}{{Ponman} et~al.}{1996}]{ponman_etal96}
{Ponman} T.~J.,  {Bourner} P.~D.~J.,  {Ebeling} H.,    {B{\"o}hringer} H.,
  1996, \mnras, 283, 690

\bibitem[\protect\citeauthoryear{{Porter}, {Somerville}, {Primack} \&
  {Johansson}}{{Porter} et~al.}{2014}]{Porter2014}
{Porter} L.~A.,  {Somerville} R.~S.,  {Primack} J.~R.,    {Johansson} P.~H.,
  2014, \mnras, 444, 942

\bibitem[\protect\citeauthoryear{{Powell}, {Bournaud}, {Chapon} \&
  {Teyssier}}{{Powell} et~al.}{2013}]{Powel2013}
{Powell} L.~C.,  {Bournaud} F.,  {Chapon} D.,    {Teyssier} R.,  2013, \mnras,
  434, 1028

\bibitem[\protect\citeauthoryear{{Prieto}, {Fernandez-Ontiveros}, {Bruzual},
  {Burkert}, {Schartmann} \& {Charlot}}{{Prieto} et~al.}{2019}]{Prieto2019}
{Prieto} M.~A.,  {Fernandez-Ontiveros} J.~A.,  {Bruzual} G.,  {Burkert} A.,
  {Schartmann} M.,    {Charlot} S.,  2019, \mnras, 485, 3264

\bibitem[\protect\citeauthoryear{{Quinn}, {Hernquist} \& {Fullagar}}{{Quinn}
  et~al.}{1993}]{Quinn1993}
{Quinn} P.~J.,  {Hernquist} L.,    {Fullagar} D.~P.,  1993, \apj, 403, 74

\bibitem[\protect\citeauthoryear{{Rodighiero}, {Brusa}, {Daddi}, {Negrello},
  {Mullaney}, {Delvecchio}, {Lutz}, {Renzini}, {Franceschini}, {Baronchelli},
  {Pozzi}, {Gruppioni}, {Strazzullo}, {Cimatti} \& {Silverman}}{{Rodighiero}
  et~al.}{2015}]{Rodighiero2015}
{Rodighiero} G.,  {Brusa} M.,  {Daddi} E.,  {Negrello} M.,  {Mullaney} J.~R.,
  {Delvecchio} I.,  {Lutz} D.,  {Renzini} A.,  {Franceschini} A.,
  {Baronchelli} I.,  {Pozzi} F.,  {Gruppioni} C.,  {Strazzullo} V.,  {Cimatti}
  A.,    {Silverman} J.,  2015, \apjl, 800, L10

\bibitem[\protect\citeauthoryear{{Roediger} \& {Br{\"u}ggen}}{{Roediger} \&
  {Br{\"u}ggen}}{2007}]{Roediger2007}
{Roediger} E.,  {Br{\"u}ggen} M.,  2007, \mnras, 380, 1399

\bibitem[\protect\citeauthoryear{{Saglia}, {Opitsch}, {Erwin}, {Thomas},
  {Beifiori}, {Fabricius}, {Mazzalay}, {Nowak}, {Rusli} \& {Bender}}{{Saglia}
  et~al.}{2016}]{Saglia2016}
{Saglia} R.~P.,  {Opitsch} M.,  {Erwin} P.,  {Thomas} J.,  {Beifiori} A.,
  {Fabricius} M.,  {Mazzalay} X.,  {Nowak} N.,  {Rusli} S.~P.,    {Bender} R.,
  2016, \apj, 818, 47

\bibitem[\protect\citeauthoryear{{Sahu}, {Graham} \& {Davis}}{{Sahu}
  et~al.}{2019}]{Sahu2019}
{Sahu} N.,  {Graham} A.~W.,    {Davis} B.~L.,  2019, \apj, 876, 155

\bibitem[\protect\citeauthoryear{{Sahu}, {Graham} \& {Davis}}{{Sahu}
  et~al.}{2022}]{Sahu2022}
{Sahu} N.,  {Graham} A.~W.,    {Davis} B.~L.,  2022, \apj, 927, 67

\bibitem[\protect\citeauthoryear{{Salpeter}}{{Salpeter}}{1964}]{Salpeter1964}
{Salpeter} E.~E.,  1964, \apj, 140, 796

\bibitem[\protect\citeauthoryear{{Sandage}}{{Sandage}}{1986}]{Sandage1986}
{Sandage} A.,  1986, \aap, 161, 89

\bibitem[\protect\citeauthoryear{{Sanders}, {Soifer}, {Elias}, {Madore},
  {Matthews}, {Neugebauer} \& {Scoville}}{{Sanders} et~al.}{1988}]{Sanders1988}
{Sanders} D.~B.,  {Soifer} B.~T.,  {Elias} J.~H.,  {Madore} B.~F.,  {Matthews}
  K.,  {Neugebauer} G.,    {Scoville} N.~Z.,  1988, \apj, 325, 74

\bibitem[\protect\citeauthoryear{{Shankar}, {Bernardi}, {Sheth}, {Ferrarese},
  {Graham}, {Savorgnan}, {Allevato}, {Marconi}, {L{\"a}sker} \&
  {Lapi}}{{Shankar} et~al.}{2016}]{Shankar2016}
{Shankar} F.,  {Bernardi} M.,  {Sheth} R.~K.,  {Ferrarese} L.,  {Graham} A.~W.,
   {Savorgnan} G.,  {Allevato} V.,  {Marconi} A.,  {L{\"a}sker} R.,    {Lapi}
  A.,  2016, \mnras, 460, 3119

\bibitem[\protect\citeauthoryear{{Silk} \& {Rees}}{{Silk} \&
  {Rees}}{1998}]{Silk1998}
{Silk} J.,  {Rees} M.~J.,  1998, \aap, 331, L1

\bibitem[\protect\citeauthoryear{{Simpson}, {Grand}, {G{\'o}mez}, {Marinacci},
  {Pakmor}, {Springel}, {Campbell} \& {Frenk}}{{Simpson}
  et~al.}{2018}]{Simpson2018}
{Simpson} C.~M.,  {Grand} R. J.~J.,  {G{\'o}mez} F.~A.,  {Marinacci} F.,
  {Pakmor} R.,  {Springel} V.,  {Campbell} D. J.~R.,    {Frenk} C.~S.,  2018,
  \mnras, 478, 548

\bibitem[\protect\citeauthoryear{{Somerville} \& {Dav{\'e}}}{{Somerville} \&
  {Dav{\'e}}}{2015}]{Somerville2015a}
{Somerville} R.~S.,  {Dav{\'e}} R.,  2015, \araa, 53, 51

\bibitem[\protect\citeauthoryear{{Somerville}, {Hopkins}, {Cox}, {Robertson} \&
  {Hernquist}}{{Somerville} et~al.}{2008}]{Somerville2008}
{Somerville} R.~S.,  {Hopkins} P.~F.,  {Cox} T.~J.,  {Robertson} B.~E.,
  {Hernquist} L.,  2008, \mnras, 391, 481

\bibitem[\protect\citeauthoryear{{Somerville}, {Popping} \&
  {Trager}}{{Somerville} et~al.}{2015}]{Somerville2015b}
{Somerville} R.~S.,  {Popping} G.,    {Trager} S.~C.,  2015, \mnras, 453, 4337

\bibitem[\protect\citeauthoryear{{Springel}, {Di Matteo} \&
  {Hernquist}}{{Springel} et~al.}{2005}]{Springel2005b}
{Springel} V.,  {Di Matteo} T.,    {Hernquist} L.,  2005, \mnras, 361, 776

\bibitem[\protect\citeauthoryear{{Springel} \& {Hernquist}}{{Springel} \&
  {Hernquist}}{2005}]{Springel2005a}
{Springel} V.,  {Hernquist} L.,  2005, \apjl, 622, L9

\bibitem[\protect\citeauthoryear{{Statler}}{{Statler}}{1988}]{Statler1988}
{Statler} T.~S.,  1988, \apj, 331, 71

\bibitem[\protect\citeauthoryear{{Steinhauser}, {Schindler} \&
  {Springel}}{{Steinhauser} et~al.}{2016}]{Steinhauser2016}
{Steinhauser} D.,  {Schindler} S.,    {Springel} V.,  2016, \aap, 591, A51

\bibitem[\protect\citeauthoryear{{Terrazas}, {Bell}, {Henriques}, {White},
  {Cattaneo} \& {Woo}}{{Terrazas} et~al.}{2016}]{Terrazas2016}
{Terrazas} B.~A.,  {Bell} E.~F.,  {Henriques} B. M.~B.,  {White} S. D.~M.,
  {Cattaneo} A.,    {Woo} J.,  2016, \apjl, 830, L12

\bibitem[\protect\citeauthoryear{{Tollet}, {Cattaneo}, {Macci{\`o}}, {Dutton}
  \& {Kang}}{{Tollet} et~al.}{2019}]{Tollet2019}
{Tollet} {\'E}.,  {Cattaneo} A.,  {Macci{\`o}} A.~V.,  {Dutton} A.~A.,
  {Kang} X.,  2019, \mnras

\bibitem[\protect\citeauthoryear{{Tollet}, {Cattaneo}, {Macci{\`o}} \&
  {Kang}}{{Tollet} et~al.}{2022}]{Tollet2022}
{Tollet} E.,  {Cattaneo} A.,  {Macci{\`o}} A.~V.,    {Kang} X.,  2022, \mnras

\bibitem[\protect\citeauthoryear{{Tollet}, {Cattaneo}, {Mamon}, {Moutard} \&
  {van den Bosch}}{{Tollet} et~al.}{2017}]{Tollet17}
{Tollet} {\'E}.,  {Cattaneo} A.,  {Mamon} G.~A.,  {Moutard} T.,    {van den
  Bosch} F.~C.,  2017, \mnras, 471, 4170

\bibitem[\protect\citeauthoryear{{Tomczak}, {Quadri}, {Tran}, {Labb{\'e}},
  {Straatman}, {Papovich}, {Glazebrook}, {Allen}, {Brammer}, {Kacprzak},
  {Kawinwanichakij}, {Kelson}, {McCarthy} \& et al.}{{Tomczak}
  et~al.}{2014}]{Tomczak2014}
{Tomczak} A.~R.,  {Quadri} R.~F.,  {Tran} K.-V.~H.,  {Labb{\'e}} I.,
  {Straatman} C.~M.~S.,  {Papovich} C.,  {Glazebrook} K.,  {Allen} R.,
  {Brammer} G.~B.,  {Kacprzak} G.~G.,  {Kawinwanichakij} L.,  {Kelson} D.~D.,
  {McCarthy} P.~J.,    et al. 2014, \apj, 783, 85

\bibitem[\protect\citeauthoryear{{Tonnesen} \& {Bryan}}{{Tonnesen} \&
  {Bryan}}{2009}]{Tonnesen2009}
{Tonnesen} S.,  {Bryan} G.~L.,  2009, \apj, 694, 789

\bibitem[\protect\citeauthoryear{{Toomre} \& {Toomre}}{{Toomre} \&
  {Toomre}}{1972}]{Toomre1972}
{Toomre} A.,  {Toomre} J.,  1972, \apj, 178, 623

\bibitem[\protect\citeauthoryear{{Valageas} \& {Silk}}{{Valageas} \&
  {Silk}}{1999}]{valageas_silk99}
{Valageas} P.,  {Silk} J.,  1999, \aap, 350, 725

\bibitem[\protect\citeauthoryear{{van den Bosch}}{{van den
  Bosch}}{1998}]{Bosch1998}
{van den Bosch} F.~C.,  1998, \apj, 507, 601

\bibitem[\protect\citeauthoryear{{Voit}}{{Voit}}{2019}]{Voit2019b}
{Voit} G.~M.,  2019, \apj, 880, 139

\bibitem[\protect\citeauthoryear{{Voit}, {Babul}, {Babyk}, {Bryan}, {Chen},
  {Donahue}, {Fielding}, {Gaspari}, {Li}, {McDonald}, {O'Shea}, {Prasad} \& et
  al.}{{Voit} et~al.}{2019}]{Voit2019c}
{Voit} M.,  {Babul} A.,  {Babyk} I.,  {Bryan} G.,  {Chen} H.-W.,  {Donahue} M.,
   {Fielding} D.,  {Gaspari} M.,  {Li} Y.,  {McDonald} M.,  {O'Shea} B.,
  {Prasad} D.,    et al. 2019, \baas, 51, 405

\bibitem[\protect\citeauthoryear{{Wang}, {Dutton}, {Stinson}, {Macci{\`o}},
  {Gutcke} \& {Kang}}{{Wang} et~al.}{2017}]{Wang2017}
{Wang} L.,  {Dutton} A.~A.,  {Stinson} G.~S.,  {Macci{\`o}} A.~V.,  {Gutcke}
  T.,    {Kang} X.,  2017, \mnras, 466, 4858

\bibitem[\protect\citeauthoryear{{Wang}, {Fazio}, {Ashby}, {Huang}, {Pahre},
  {Smith}, {Willner}, {Forrest}, {Pipher} \& {Surace}}{{Wang}
  et~al.}{2004}]{Wang2004}
{Wang} Z.,  {Fazio} G.~G.,  {Ashby} M.~L.~N.,  {Huang} J.~S.,  {Pahre} M.~A.,
  {Smith} H.~A.,  {Willner} S.~P.,  {Forrest} W.~J.,  {Pipher} J.~L.,
  {Surace} J.~A.,  2004, \apjs, 154, 193

\bibitem[\protect\citeauthoryear{{Weinmann}, {van den Bosch}, {Yang}, {Mo},
  {Croton} \& {Moore}}{{Weinmann} et~al.}{2006}]{Weinmann2006}
{Weinmann} S.~M.,  {van den Bosch} F.~C.,  {Yang} X.,  {Mo} H.~J.,  {Croton}
  D.~J.,    {Moore} B.,  2006, \mnras, 372, 1161

\bibitem[\protect\citeauthoryear{{Wetzel}, {Tinker}, {Conroy} \& {van den
  Bosch}}{{Wetzel} et~al.}{2013}]{Wetzel2013}
{Wetzel} A.~R.,  {Tinker} J.~L.,  {Conroy} C.,    {van den Bosch} F.~C.,  2013,
  \mnras, 432, 336

\bibitem[\protect\citeauthoryear{{Whitaker}, {Franx}, {Bezanson}, {Brammer},
  {van Dokkum}, {Kriek}, {Labb{\'e}}, {Leja}, {Momcheva}, {Nelson}, {Rigby},
  {Rix}, {Skelton}, {van der Wel} \& {Wuyts}}{{Whitaker}
  et~al.}{2015}]{Whitaker2015}
{Whitaker} K.~E.,  {Franx} M.,  {Bezanson} R.,  {Brammer} G.~B.,  {van Dokkum}
  P.~G.,  {Kriek} M.~T.,  {Labb{\'e}} I.,  {Leja} J.,  {Momcheva} I.~G.,
  {Nelson} E.~J.,  {Rigby} J.~R.,  {Rix} H.-W.,  {Skelton} R.~E.,  {van der
  Wel} A.,    {Wuyts} S.,  2015, \apjl, 811, L12

\bibitem[\protect\citeauthoryear{{White} \& {Frenk}}{{White} \&
  {Frenk}}{1991}]{White1991}
{White} S. D.~M.,  {Frenk} C.~S.,  1991, \apj, 379, 52

\bibitem[\protect\citeauthoryear{{Wisotzki}, {Kuhlbrodt} \&
  {Jahnke}}{{Wisotzki} et~al.}{2001}]{wisotzki_etal01}
{Wisotzki} L.,  {Kuhlbrodt} B.,    {Jahnke} K.,  2001, in {M{\'a}rquez} I.,
  {Masegosa} J.,  {del Olmo} A.,  {Lara} L.,  {Garc{\'\i}a} E.,   {Molina} J.,
  eds, QSO Hosts and Their Environments {The luminosity function of QSO host
  galaxies}.
p.~83

\bibitem[\protect\citeauthoryear{{Woods}, {Agarwal}, {Bromm}, {Bunker}, {Chen},
  {Chon}, {Ferrara}, {Glover}, {Haemmerl{\'e}}, {Haiman}, {Hartwig}, {Heger} \&
  et al.}{{Woods} et~al.}{2019}]{Woods2019}
{Woods} T.~E.,  {Agarwal} B.,  {Bromm} V.,  {Bunker} A.,  {Chen} K.-J.,  {Chon}
  S.,  {Ferrara} A.,  {Glover} S. C.~O.,  {Haemmerl{\'e}} L.,  {Haiman} Z.,
  {Hartwig} T.,  {Heger} A.,    et al. 2019, \pasa, 36, e027

\bibitem[\protect\citeauthoryear{{Wu}, {Wang}, {Fan}, {Yi}, {Zuo}, {Bian},
  {Jiang}, {McGreer}, {Wang}, {Yang}, {Yang}, {Thompson} \& {Beletsky}}{{Wu}
  et~al.}{2015}]{Wu2015}
{Wu} X.-B.,  {Wang} F.,  {Fan} X.,  {Yi} W.,  {Zuo} W.,  {Bian} F.,  {Jiang}
  L.,  {McGreer} I.~D.,  {Wang} R.,  {Yang} J.,  {Yang} Q.,  {Thompson} D.,
  {Beletsky} Y.,  2015, \nat, 518, 512

\bibitem[\protect\citeauthoryear{{Wuyts}, {F{\"o}rster Schreiber}, {van der
  Wel}, {Magnelli}, {Guo}, {Genzel}, {Lutz}, {Aussel}, {Barro}, {Berta} \& et
  al.}{{Wuyts} et~al.}{2011}]{Wuyts2011}
{Wuyts} S.,  {F{\"o}rster Schreiber} N.~M.,  {van der Wel} A.,  {Magnelli} B.,
  {Guo} Y.,  {Genzel} R.,  {Lutz} D.,  {Aussel} H.,  {Barro} G.,  {Berta} S.,
   et al. 2011, \apj, 742, 96

\bibitem[\protect\citeauthoryear{{Yang}, {Mo} \& {van den Bosch}}{{Yang}
  et~al.}{2009}]{Yang2009}
{Yang} X.,  {Mo} H.~J.,    {van den Bosch} F.~C.,  2009, \apj, 693, 830

\end{thebibliography}

\end{document}